\documentclass[dvipsnames]{aa}

\usepackage[utf8]{inputenc}
\usepackage[T1]{fontenc}
\usepackage[dvipsnames]{xcolor}
\usepackage{txfonts}
\usepackage{graphicx}

\bibpunct{(}{)}{;}{a}{}{,} % to follow the A&A style

\newcommand{\grvs}{G$_{\rm{RVS}}$}	
\newcommand{\teff}{T$_{\rm{eff}}$}

\begin{document}
\titlerunning{Stellar rotation and binaries in open clusters with Gaia DR3}
\title{Stellar rotation and binaries in open clusters with Gaia DR3}
\author{E.~Pancino\inst{\ref{oaa}},
        E. Reggiani\inst{\ref{oaa}, \ref{unifi}},
        S.~Marinoni\inst{\ref{oar}},
        P.~M.~Marrese\inst{\ref{oar}},
        D.~Alvarez Garay\inst{\ref{oaa}},
        A.~Avdeeva\inst{\ref{oaa}},
        M.~Echeveste\inst{\ref{oaa}},
        E.~Leitinger\inst{\ref{oaa}, \ref{unibo}}
        S.~Nedhath\inst{\ref{oaa}, \ref{unifi}},
        S.~Rani\inst{\ref{oaa}},
        N.~Sanna\inst{\ref{oaa}}, 
        S.~Saracino\inst{\ref{oaa}},
        L.~Steinbauer\inst{\ref{oaa}, \ref{unifi}},
        A.~Turchi\inst{\ref{oaa}},
        V.~V.~Jadhav\inst{\ref{ai}},
        S.~Kamann\inst{\ref{liv}},
        M.~Rainer\inst{\ref{oato}}}

\institute{INAF - Osservatorio Astrofisico di Arcetri, Largo E. Fermi 5, 50125 Firenze, Italy\label{oaa}
\and Dipartimento di Fisica e Astronomia, Universit{\`a} di Firenze,  Via G. Sansone 1, 50019 Sesto Fiorentino, FI, Italy\label{unifi}
\and INAF -- Osservatorio Astronomico di Roma, Via Frascati 33, 00040, Monte Porzio Catone, Roma, Italy\label{oar}
\and Dipartimento di Fisica e Astronomia, Universit{\`a} degli Studi di Bologna, Via Gobetti 93/2, 40129 Bologna, Italy\label{unibo}
\and Astronomical Institute, Faculty of Mathematics and Physics, Charles University, V Hole\v{s}ovi\v{c}k\'{a}ch 2, CZ-180 00 Praha 8, Czech Republic\label{ai}
\and Astrophysics Research Institute, Liverpool John Moores University, IC2 Liverpool Science Park, 146 Brownlow Hill, Liverpool, L3 5RF, United Kingdom\label{liv}
\and INAF -- Osservatorio Astrofisico di Torino, Strada Osservatorio 20, 10025, Pino Torinese, TO, Italy\label{oato}}
\authorrunning{E. Pancino et al.}   
\date{Received: ---}
 
\abstract
{Stellar rotation is a fundamental ingredient in shaping the evolution of stars and it can also be used to trace past stellar interactions. Yet, systematic studies of stellar rotation in large samples of stars belonging to different populations have only recently been made possible, thanks to spectroscopic surveys.}
{We profit from the catalogue of rotational broadening and rotation periods released with {\em Gaia} DR3. We focus on open clusters to study the rotational behaviour of several interesting populations including, among others, blue stragglers and extended main sequence turnoffs (eMSTO).}
{We use literature lists of almost a million member stars in several thousand open clusters in the Milky Way. We collect properties of stars and clusters from large surveys, including {\em Gaia}, and from various literature sources. We include a comprehensive collection of known variables and binary stars from various databases. We manually select (exotic) stellar populations from the color-magnitude diagrams of individual clusters and study their rotational properties.}
{Our catalogue contains more than 44\,000 rotationally characterised stars, almost 57\,000 variables (excluding binaries) and more than 22\,000 binary stars. We find several interesting results, including a few hundred new blue stragglers, several fast rotating red giants, and we increase the number of clusters with an eMSTO to 96. We discover that most clusters more massive than 10$^3$M$_{\rm{\odot}}$ display an eMSTO. We present a new parametrization of the number of blue stragglers as a function of cluster mass and age. We find that the percentage of binary stars in the equal-mass binary sequence and in the main sequence are similar.}
{We present the first large-scale statistical exploration of stellar rotation in open clusters, which already yielded new interesting results and which can be used as the basis for several detailed follow-up studies.}


\maketitle

%%%%%%%%%%%%%%%%%%%%%%%%%%%%%%%%%%%%%%%%%%%%%%%%%%%%%%%%%%%%%%%%%%%%%%%%%%%%%%%%%%%%%%

\section{Introduction}
\label{sec:intro}

Stellar rotation rates are a necessary ingredient to understand how stars form, their evolutionary pathways, and their surface chemistry, among others. Stellar rotation significantly affects internal processes, such as mixing and angular momentum transport. For instance, fast-rotating stars may exhibit enhanced chemical mixing, influencing isotopic ratios like [C/O] and [N/O], which provide insights into stellar evolution stages \citep{zahn92,maeder00}. Fast rotating massive stars were in fact key to explain the primary nitrogen production observed in Milky Way stellar abundance ratios \citep{meynet02}. Stellar rotation is also important in the context of the evolution of binary star systems. Interaction in close binaries can result in tidal locking and synchronization, while mass transfer episodes can also transfer considerable angular momentum \citep{hut81,soberman97,hurley02}. Such interactions can lead to significant changes in the rotation rates and in the subsequent evolution of the stars involved. Stellar rotation rates can also influence the mass loss and wind properties in massive stars \citep{smith14}. Finally, stellar rotation evolves naturally during the life of stars, particularly low-mass stars, because of magnetic braking, opening the possibility for stellar dating techniques such as gyrochronology \citep{kraft67,skumanich72,barnes03}. Systematic measurements of stellar rotation rates across the whole parameter space of stellar properties and environments require exceptional observational efforts. In particular, photometric monitoring and asteroseismic surveys (see App.~\ref{sec:var} for references) have provided large samples of rotational period measurements, while large spectroscopic surveys (see Sect.~\ref{sec:rot} for references) provided spectral line broadening measurements for even larger samples. 

Here, we focus on open clusters, which offer the unique opportunity of having accurate age and distance estimates for hundreds of thousands of member stars. More specifically, the {\em Gaia} mission \citep{gaia} has drastically improved our samples of open clusters and member stars, among many other results. In particular, the third {\em Gaia} data release \citep[hereafter DR3,][]{gdr3}, provides parameters related to the rotational characterization of stars for the first time (see Sect.~\ref{sec:rot} for more details and references). Although these measurements are expected to improve in future releases and are subject to some limitations, the sheer amount of characterised stars opens the possibility to carry out an exploratory study, a statistical overview of open cluster stars across the color-magnitude diagram (CMD), which can provide a useful starting point for more detailed investigations.

% --------------------------------------------
\begin{figure}
    \centering
	\includegraphics[width=\columnwidth]{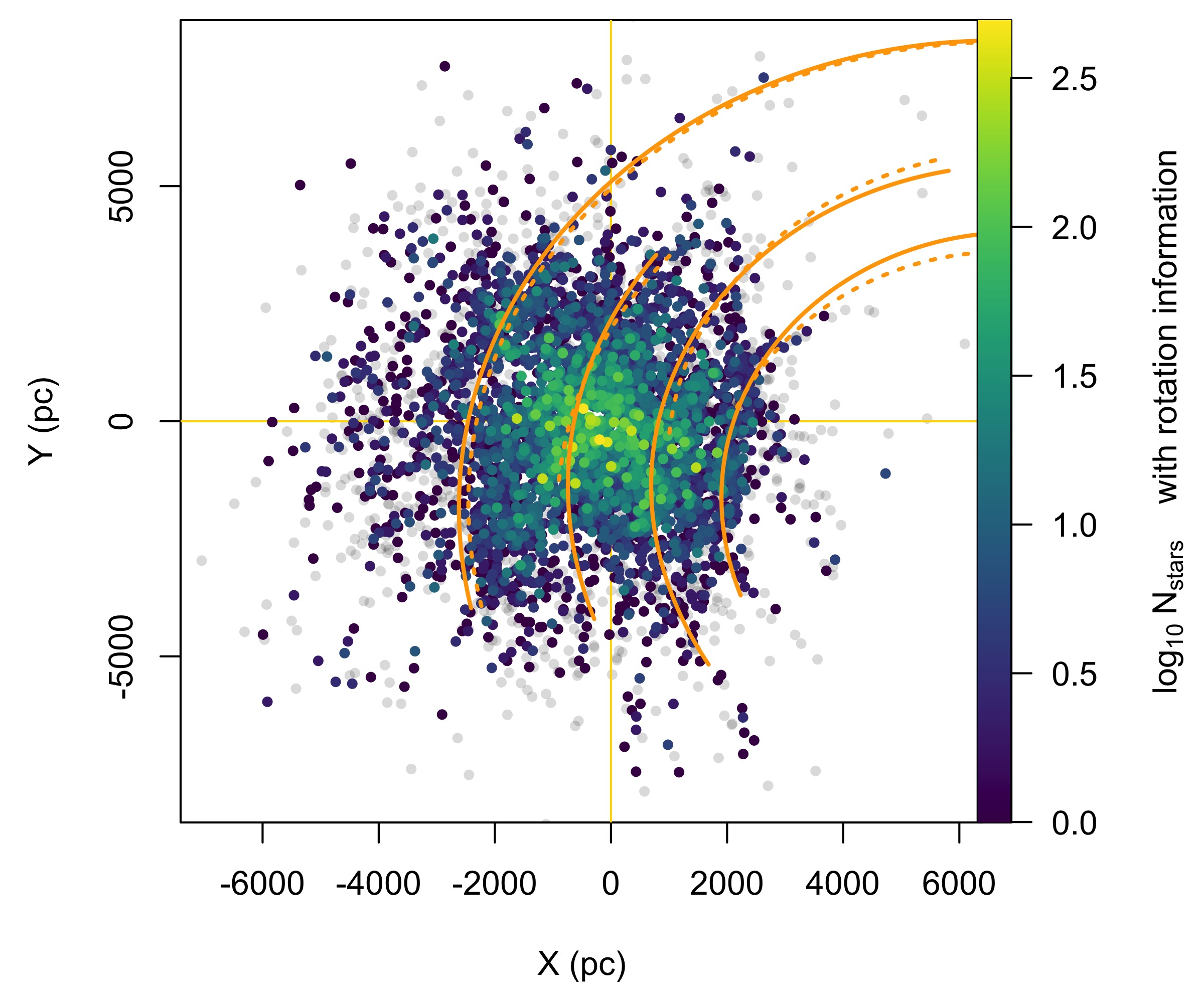}
	\vspace{-0.5cm}
    \caption{The studied clusters in Galactic cartesian coordinates. All clusters are plotted in grey in the background. Clusters are colored based on the number of stars with rotation information. The color scale saturates at about 400 for clarity, but several clusters have up to a few thousands of members with rotation information. The spiral arms of the Milky Way, as modelled by \citet{castro21} and \citet{reid14}, are plotted as solid and dashed orange lines, respectively. }
   \label{fig:ocs}
\end{figure}
% --------------------------------------------

The paper is organised as follows: our sample is presented in Sect.~\ref{sec:data}; we explore the rotational properties of different stellar populations, including exotic ones, in Sect.~\ref{sec:res}; we summarise our main results and present our conclusions in Sect.~\ref{sec:conc}. Further details on the characterization of member stars and clusters, as well as additional figures, are presented in the Appendices. 

% ------------------------------------------------
\begin{table}
	\centering
	\caption{Catalogue of clusters for which at least one member star has a {\em Gaia} DR3 {\tt vbroad}, {\tt vsini}, or {\tt prot} estimate, with selected properties. See Section~\ref{sec:ocs} for more details. The Table is available electronically, where reported quantities are accompanied by their uncertainties.}
	\label{tab:ocs}
	\begin{tabular}{lll} 
		\hline
		Quantity & Units & Description \\
		\hline
		Cluster & & Adopted cluster name\\
		AltNames & & Other cluster names\\
		RA$_0$ & deg & Central Right Ascension \\
		Dec$_0$ & deg & Central Declination \\
		pmRA$^*$ & mas/yr & Mean proper motion in RA \\
		pmDec & mas/yr & Mean proper motion in Dec \\
		$\varpi$ & mas & Mean parallax \\
		X & pc & X in Galactic cartesian coordinates \\
		Y & pc & Y in Galactic cartesian coordinates \\		
		Z & pc & Z in Galactic cartesian coordinates\\
		Dist & pc & Distance from the cluster \\
		R$_{\rm{GC}}$ & pc & Galactocentric distance \\
        R$_{50}$ & deg & Radius enclosing half members \\
        R$_{\rm J}$ & deg & Jacobi radius \\
		n$_{\rm{mem}}$ & & Number of cluster members \\
		sourceAstro & & Literature source of astrometry \\
	    RV & km\,s$^{-1}$ & Mean RV for the cluster \\
        n$_{\rm{RV}}$ & & Stars used for RV \\
        SourceRV & & Literature source for RV \\
	    $[$Fe/H$]$ & dex & Mean iron metallicity of the cluster \\
        n$_{\rm{[Fe/H]}}$ & & Stars used for [Fe/H] \\
		SourceFeH & & Literature source of [Fe/H] \\
		log(age) & yr & Logarithmic age of the cluster \\
		SourceAge & & Literature source of log(age) \\
		A$_{\rm V}$ & mag & Absorption in the V band \\
	        mass & M$_{\odot}$ & Cluster mass \\
        has\_rot & & $\geq$1 star with rotation information\\
		\hline
	\end{tabular}
\end{table}
% ------------------------------------------------

%%%%%%%%%%%%%%%%%%%%%%%%%%%%%%%%%%%%%%%%%%%%%%%%%%

\section{Sample selection and characterization}
\label{sec:data}

\subsection{Open clusters list}
\label{sec:ocmaster}

Since the advent of the {\em Gaia} space mission \citep{gaia}, the census of existing open clusters has undergone a revolution \citep[see Fig.~1 and Table~1 by][]{perren23}, with significant increases in sample size and data quality compared to previous efforts \citep{kharchenko13,dias02}. In this panorama, the most comprehensive effort was conducted by \citet{hunt21,hunt23,hunt24}, based on {\em Gaia} DR3 astrometry and photometry \citep{gdr3}. We based our work mostly on their catalog, integrated with literature data as described in the appendices. There are 5647 open clusters ({\tt kind\,=\,"o"}) listed by \citet{hunt24}. However, the authors note that $\simeq$\,10--20\% of the well-known clusters in the literature are not found with their method. For this reason, we identified 51 missing clusters possessing good {\em Gaia} data, which were added to our cluster master list (see App.~\ref{app:sample} for details). 

Fig.~\ref{fig:ocs} shows the spatial distribution of our sample clusters. We also critically compiled cluster properties from the literature and re-determined some of them, using information from {\em Gaia} DR3 and from spectroscopic surveys, as described in full detail in App.~\ref{app:sample}. The final list of clusters with their selected properties can be found in Table~\ref{tab:ocs}, which is available electronically.

%%%%%%%%%%%%%%%%%%%%%%%%%%%%%%%%%%%%%%%%%%%%%%%%%%

% ------------------------------------------------
\begin{table}
	\centering
	\caption{Catalogue of member stars in the full sample, along with some fundamental properties obtained from {\em Gaia} DR3 and other literature sources (see App.~\ref{app:params} for more details). Each property is accompanied by its uncertainty in the full electronic version of the Table.}
	\label{tab:stars}
	\begin{tabular}{lll} 
		\hline
		Column & Units & Description \\
		\hline
		source\_id & & {\em Gaia} DR3 identifier \\
		Cluster & & Adopted cluster name \\
		RA & deg & Right Ascension \\
		Dec & deg & Declination \\
		$\mu_{\rm{RA}}^*$ & mas\,yr$^{-1}$ & RA proper motion \\
		$\mu_{\rm{Dec}}$ & mas\,yr$^{-1}$ & Dec proper motion \\
		$\varpi$ & mas & Parallax \\
		G & mag & G-band magnitude \\
		G$_{\rm{BP}}$ & mag & BP magnitude \\
		G$_{\rm{RP}}$ & mag & RP magnitude \\
		E(G$_{\rm{BP}}$--G$_{\rm{RP}}$) & mag & Reddening \\
		A$_{\rm{G}}$ & mag & Absorption \\
		RV & km\,s$^{-1}$ & Radial velocity \\
		RV\_mem & & Radial velocity membership \\
		  & & (yes, no, maybe, or empty) \\
		{\tt vbroad} & km\,s$^{-1}$ & Rotational broadening RVS \\
        {\tt vsini} & km\,s$^{-1}$ & Rotational broadening ESP-HS \\
		{\tt prot} & days & Rotation period \\
		A$_{\rm{max}}$ & mag & Photospheric activity \\
		\teff & K & Effective temperature \\
		log$g$ & dex & Surface gravity \\
        $[$Fe/H] & dex & Iron metallicity \\
		M & M$_{\rm{\odot}}$ & Mass \\
		R & R$_{\rm{\odot}}$ & Radius \\
		flag\_var & & variability flag (Tab.~\ref{tab:varinfo}) \\
                  & & ({\tt VAR}, {\tt VAR:}, or empty)\\
		flag\_bin & & binarity flag (Tab.~\ref{tab:bininfo})\\
                  & & ({\tt BIN}, {\tt BIN:}, or empty) \\
		%star\_type & & See Table~\ref{tab:varinfo} \\
		star\_sample & & See Sect.~\ref{sec:res} \\
		\hline
	\end{tabular}
\end{table}
% ------------------------------------------------

% --------------------------------------------
\begin{figure*}
    \centering
	\includegraphics[width=\textwidth]{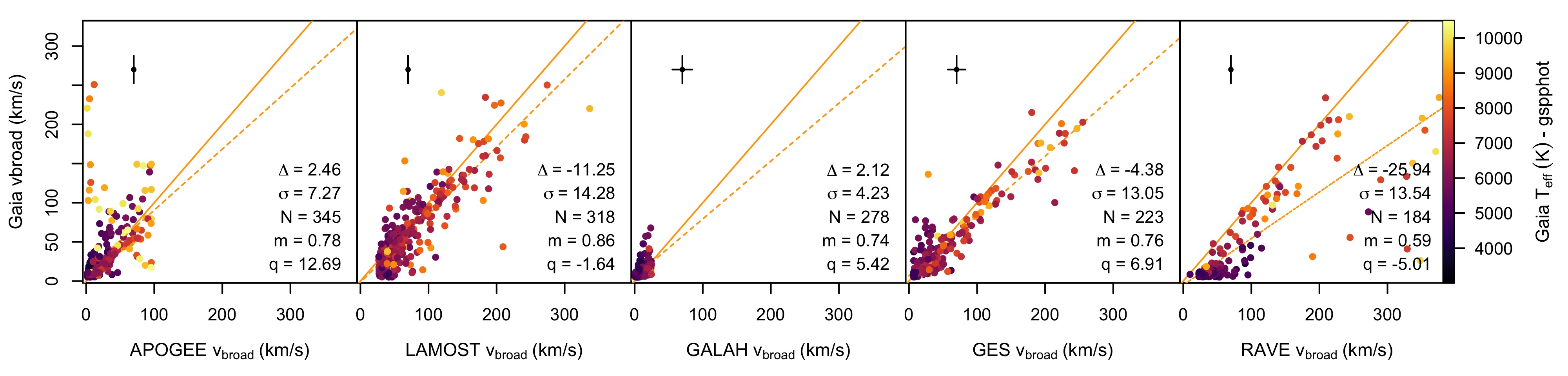}
	\includegraphics[width=\textwidth]{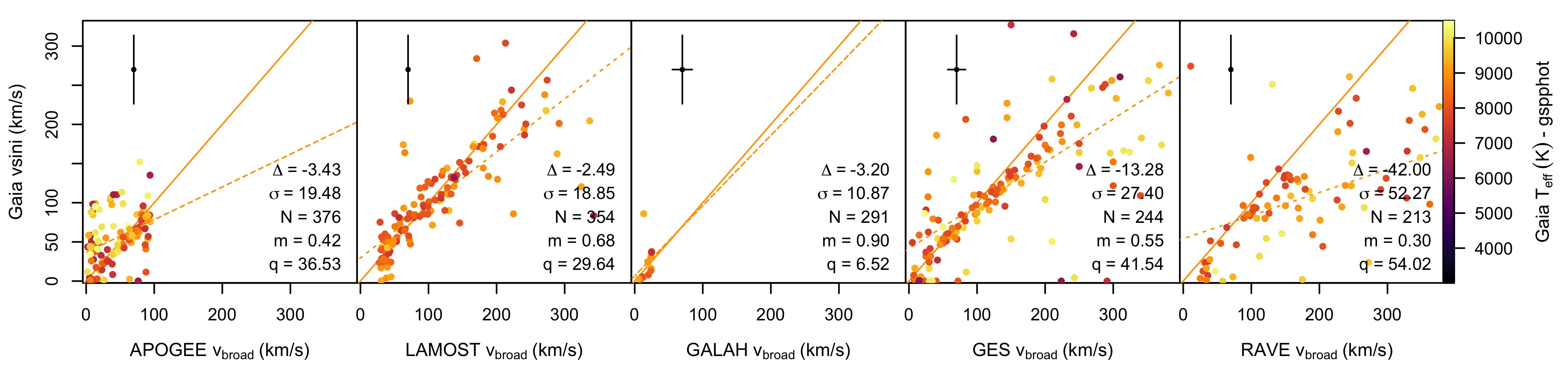}
	\vspace{-0.3cm}
    \caption{Comparison of the {\em Gaia} DR3 {\tt vbroad} (top panels) and {\tt vsini} (bottom panels) estimates with those of five spectroscopic surveys. The one-to-one agreement is reported in each panel as a solid orange line, while a linear fit to the data is reported as a dashed line. Annotated in each panel there are: the median difference ($\Delta$), the median absolute deviation (MAD, $\sigma$), the number of stars (N), the angular coefficient (m) and the intercept (q) of the linear fit. Points are colored according to T$_{\rm{eff}}$.}
   \label{fig:surveys}
\end{figure*}
% --------------------------------------------

% --------------------------------------------
\begin{figure}
    \centering
	\includegraphics[width=\columnwidth]{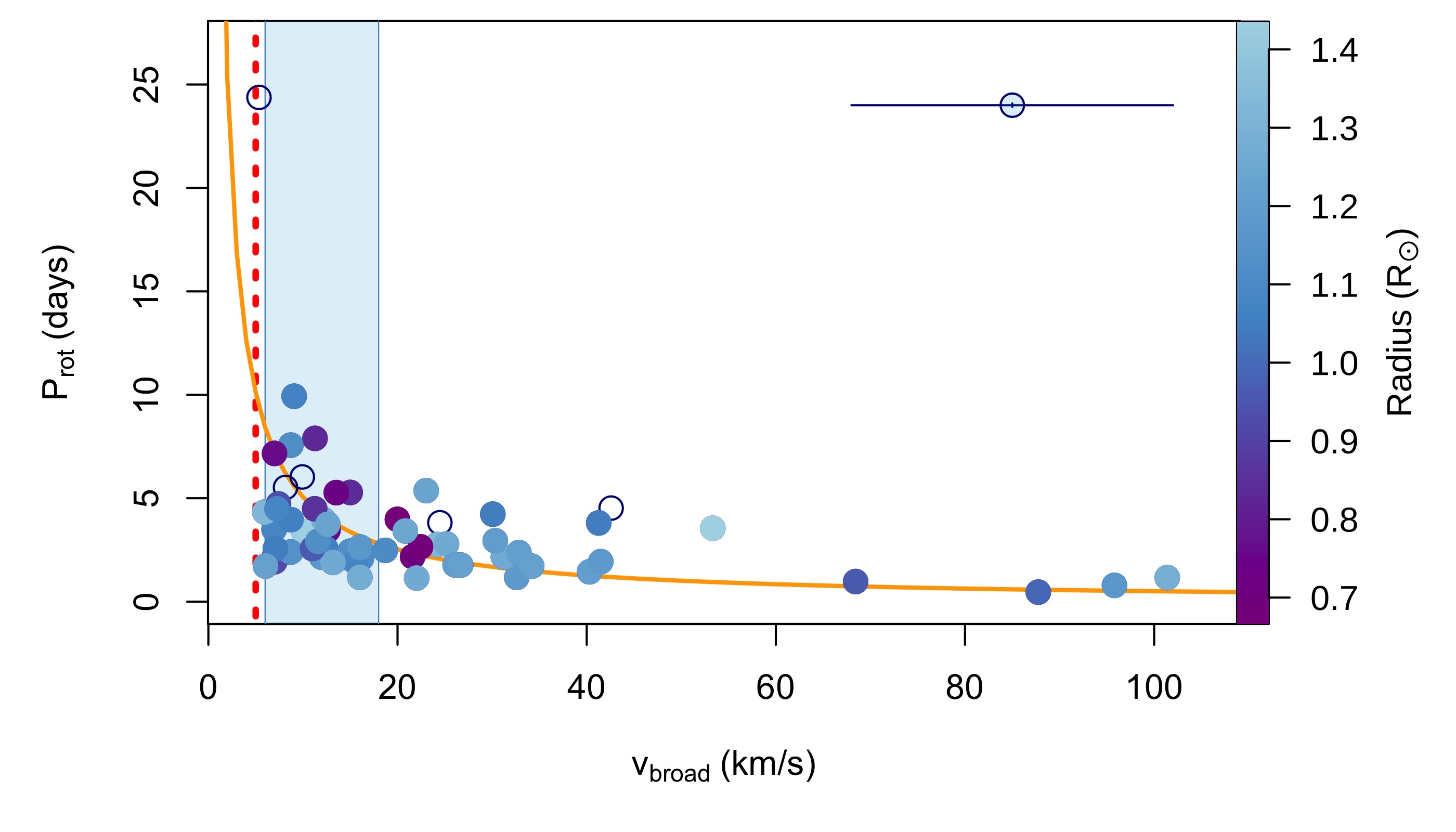}
	\vspace{-0.5cm}
    \caption{Relation between {\tt prot} and {\tt vbroad/vsini} for the 65 dwarf stars which have both measurements. The region where {\tt vbroad} starts to be overestimated is shaded in light blue, the filtering at 5\,km\,s$^{-1}$ is represented by a red dashed line, and the expected relationship for a 1\,R$_{\odot}$ star with sin\,$i$\,=\,1 is plotted in orange. The typical (median) errorbar of the 65 stars is plotted in the upper right corner. Stars are colored according to their radius (empty symbols indicate stars without a radius estimate).}
   \label{fig:prot}
\end{figure}
% --------------------------------------------

\subsection{Member stars list}
\label{sec:starmaster}

The list of member stars published by \citet{hunt24} contains more than 700\,000 stars belonging to open clusters, out of a total of almost 1.3 million stars. After a series of extensive comparisons with other member lists in the literature, we found several advantages in using the \citet{hunt24} list: their work is based on {\em Gaia} DR3 while several other lists are based on DR2; the CMDs appear cleaner and better defined; stars belonging to the tidal tails of clusters are included; and the magnitude limit is deeper than several previous searches, reaching G\,$\simeq$\,20 instead of 18\,mag. To integrate the \citet{hunt24} members list with the 51 missing clusters mentioned in Sect.~\ref{sec:ocmaster} and to characterise their properties, we relied on our own critically compiled collection of literature sources, as discussed in details in App.~\ref{app:params}. We also performed a comprehensive literature search to identify known variable and binary stars, as described in details in App.~\ref{app:varbin}.

%%%%%%%%%%%%%%%%%%%%%%%%%%%%%%%%%%%%%%%%%%%%%%%%%%

\subsection{Information on stellar rotation from {\em Gaia DR3}}
\label{sec:rot}

The {\em Gaia} DR3 catalogue provides three means of characterizing stellar rotation. The first is the rotational broadening measured on the Radial Velocity Spectrometer (RVS) spectra \citep{gaia,cropper18,katz22}; we briefly refer to this measurement as {\tt vbroad}. More details about its strengths and limitations can be found in App.~\ref{sec:vbroad} and in \citet{fremat22}.  
We complemented the {\tt vbroad} measurements with the {\tt vsini\_esphs} ones, hereafter {\tt vsini}, obtained for hot stars from a combination of RVS spectra and spectrophotometry and described in more details in App.~\ref{sec:vsini}. These measurements are more scattered than the {\tt vbroad} ones so we will mostly rely on {\tt vbroad} in the following. Additionally, we used the rotational periods ({\tt best\_rotation\_period}) obtained from the characterization of spotted rotating stars, hereafter {\tt prot}, which are described in more details in App.~\ref{sec:prot} and in \citet{lanzafame18} and \citet{distefano22}. 

A comparison of {\tt vbroad} and {\tt vsini} with the values provided in the main spectroscopic surveys is presented in Fig.~\ref{fig:surveys}. In particular, we compare with the line broadening measured by APOGEE DR17 \citep{abdurrouf22}, LAMOST DR10\footnote{\url{http://lamost.org/dr10/}}, GALAH DR4 \citep{galahdr4}, Gaia-ESO DR5.1 \citep{gilmore12,randich22}, and RAVE DR6 \citep{steinmetz20}. As can be seen, the overall agreement is satisfactory, considering the large uncertainties of this type of measurement and the validity limits of each survey. The comparison confirms the tendency of {\em Gaia} measurements to be underestimated for stars hotter than about 7500\,K (see especially the two rightmost panels of Fig.~\ref{fig:surveys}). We also stress here that there appear to be residual biases of up to $\simeq$50\,km\,s$^{-1}$ for stars hotter than $\simeq$7000\,K as the $G$ magnitude becomes fainter \citep[see Fig.~14 by][]{fremat22}. As an overall sanity check, we compared {\tt prot} with {\tt vbroad} (complemented by {\tt vsini}) for the 65 stars in our sample that have both measurements. As can be seen from Fig.~\ref{fig:prot}, the two estimates roughly follow the expected trend and are broadly consistent with each other, which is more than satisfactory, especially when considering the uncertainties of the {\tt vbroad} and {\tt vsini} determination process.

%%%%%%%%%%%%%%%%%%%%%%%%%%%%%%%%%%%%%%%%%%%%%%%%%%

% --------------------------------------------
\begin{figure}
    \centering
    	\includegraphics[width=\columnwidth]{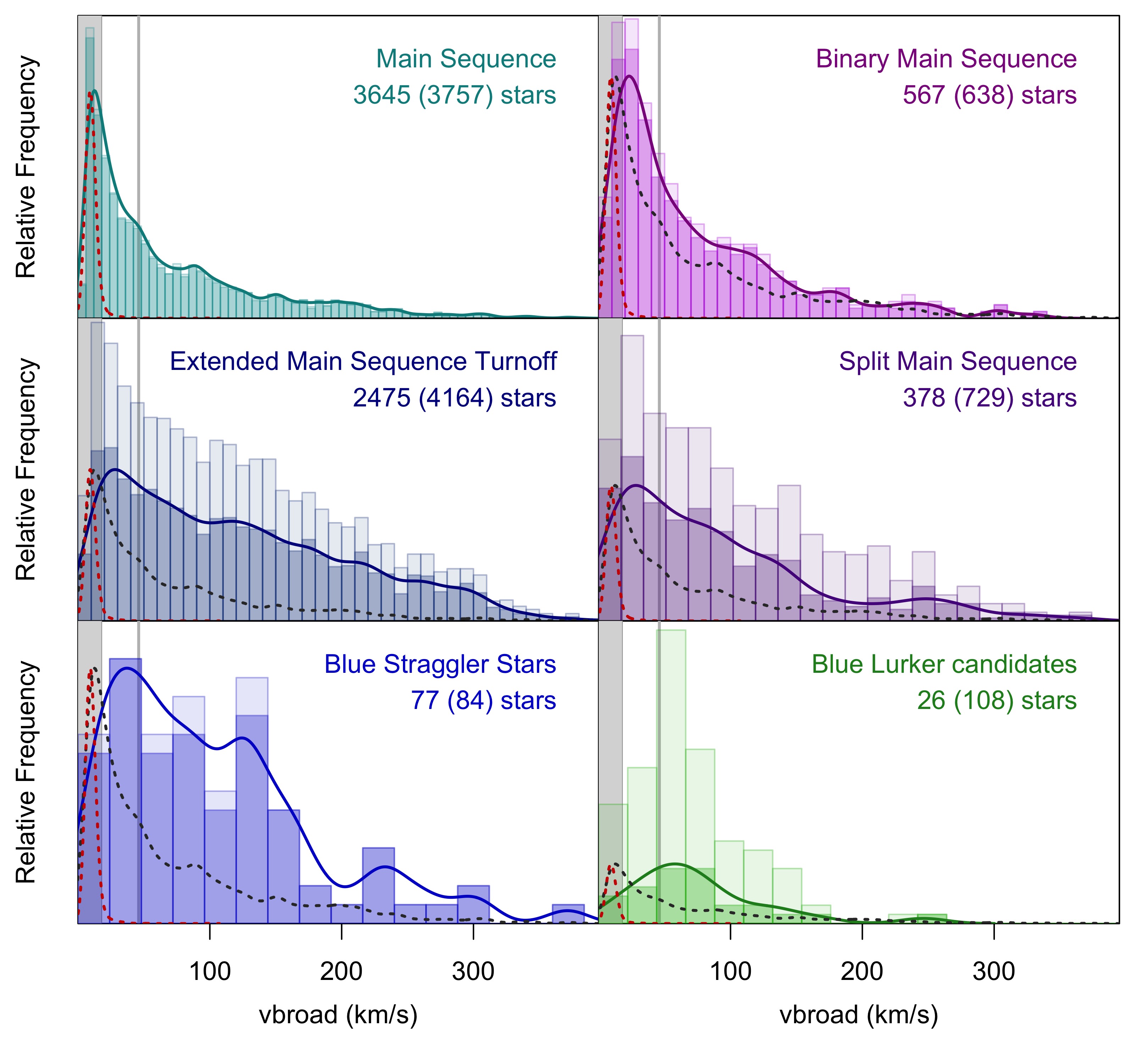}
    \vspace{-0.3cm}
    \caption{Distribution of {\tt vbroad} of dwarf samples analyzed in Sect.~\ref{sec:res}. 
    In each panel, the generalised histograms of the MS and RGB samples are reported as a grey and a red dashed line, respectively. Lighter shaded histograms include also less certain sample members. Darker histograms show only the best members (sample sizes in parenthesis). The grey shaded area ({\tt vbroad}\,$<$\,20\,km\,s$^{-1}$) in each panel represents the region where {\tt vbroad} starts to be overestimated. The vertical grey line at 46\,km\,s$^{-1}$ represents twice the RVS resolution. Given the diversity of sample sizes, the vertical axes have different (linear) scales.}
   \label{fig:histbinDW}
\end{figure}
% --------------------------------------------

\section{Stellar rotation across the CMD}
\label{sec:res}

To study the rotational properties of our sample stars, we subdivided them in different samples, indicated in the {\tt star\_sample} field in Table~\ref{tab:stars}. The selections have been done semi-manually, using strictly the data distribution on the CMD of each individual cluster, as further described below. For this reason, we focused on clusters with relatively low reddening (and apparent differential reddening $\lesssim$0.5\,mag) and with a sufficient number of stars ($\gtrsim$10) in the relevant sequences. We used solar-scaled, solar metallicity BaSTI\footnote{\url{http://basti-iac.oa-abruzzo.inaf.it/isocs.html}} isochrones and solar metallicity ZAMS \citep{hidalgo18}, with the appropriate cluster parameters from Table~\ref{tab:ocs}, exclusively to guide the eye and to double-check the cluster parameters. Even if the literature parameters are mostly obtained with PARSEC isochrones, the two sets are comparable for the purpose of the present qualitative study \citep[see][for comparisons]{hidalgo18,nguyen22,boin26}. To provide an overview of the data in hand, we show in App.~\ref{app:rot} a few CMDs and Kiel diagrams showing the position, properties, and fraction of stars with rotational properties. In the following sections we describe and discuss our results, separately for each of the selected samples.

% --------------------------------------------
\begin{figure}
    \centering
    	\includegraphics[width=\columnwidth]{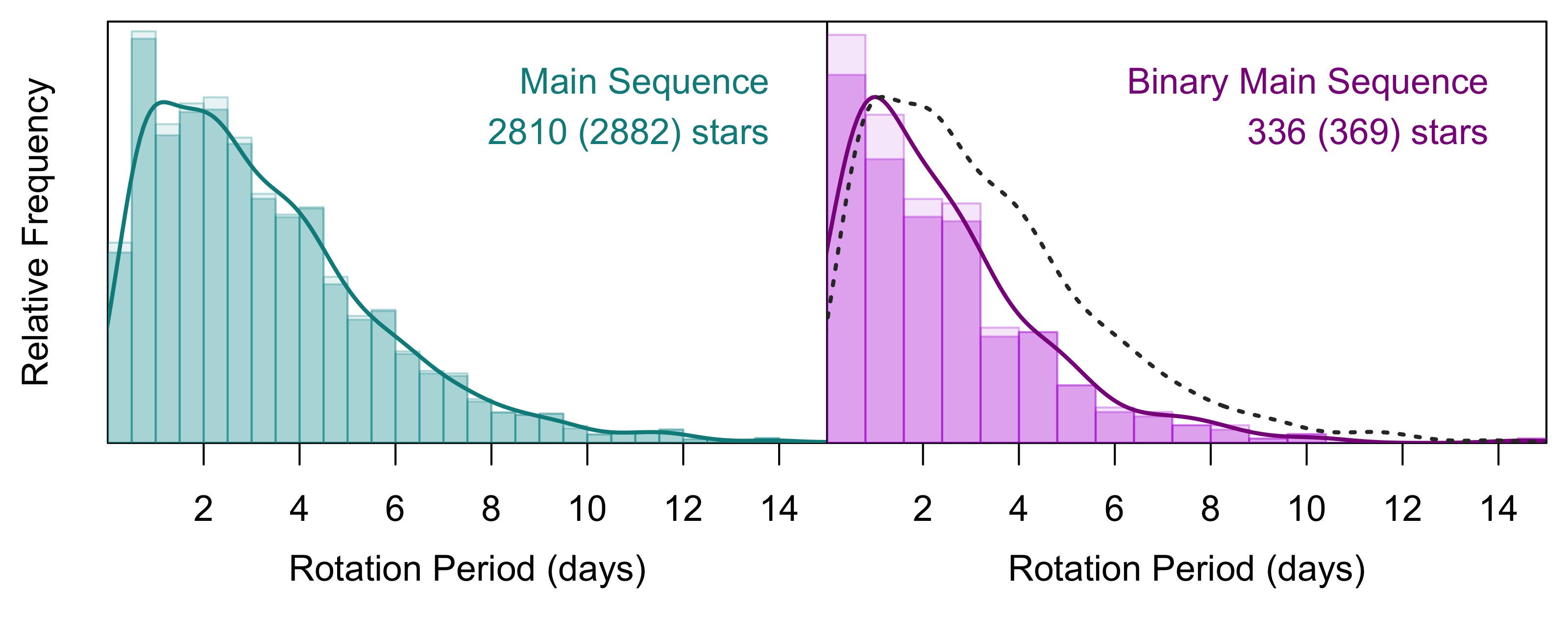}
    \vspace{-0.3cm}
    \caption{Similar to Fig.~\ref{fig:histbinDW}, but for the rotation periods, which are only available for the MS and binMS samples.}
   \label{fig:histbinPR}
\end{figure}
% --------------------------------------------

%%%%%%%%%%%%%%%%%%%%%%%%%%%%%%%%%%%%%%%%%%%%%%%%%%

\subsection{Main and binary sequence}
\label{sec:ms}
\label{sec:binms}

Main sequence (MS) stars were selected as being within 3\,MAD (median absolute deviation) from a smooth-spline fit to the MS ridge line. 
The binary MS sample (hereafter binMS) was selected in a similar way, by shifting the spline fit up by 0.75\,mag\footnote{This corresponds to twice the flux of an MS star.}. In clusters with slightly higher differential reddening (but always $\lesssim$0.5\,mag), we used the midpoint between the two fits to separate MS and binMS. Thus, our binMS sample is mostly composed of binaries with similar masses (q\,$\gtrsim$\,0.5--0.6), but this varies depending on how tight the MS of each cluster is. Examples of sample selections are presented in App.~\ref{sec:binfrac}. We further excluded more than 100 abnormally fast rotators, which we flag as candidate blue lurkers (BL), as discussed in detail in App.~\ref{sec:bl}.

% --------------------------------------------
\begin{figure*}
    \centering
	\includegraphics[width=\textwidth]{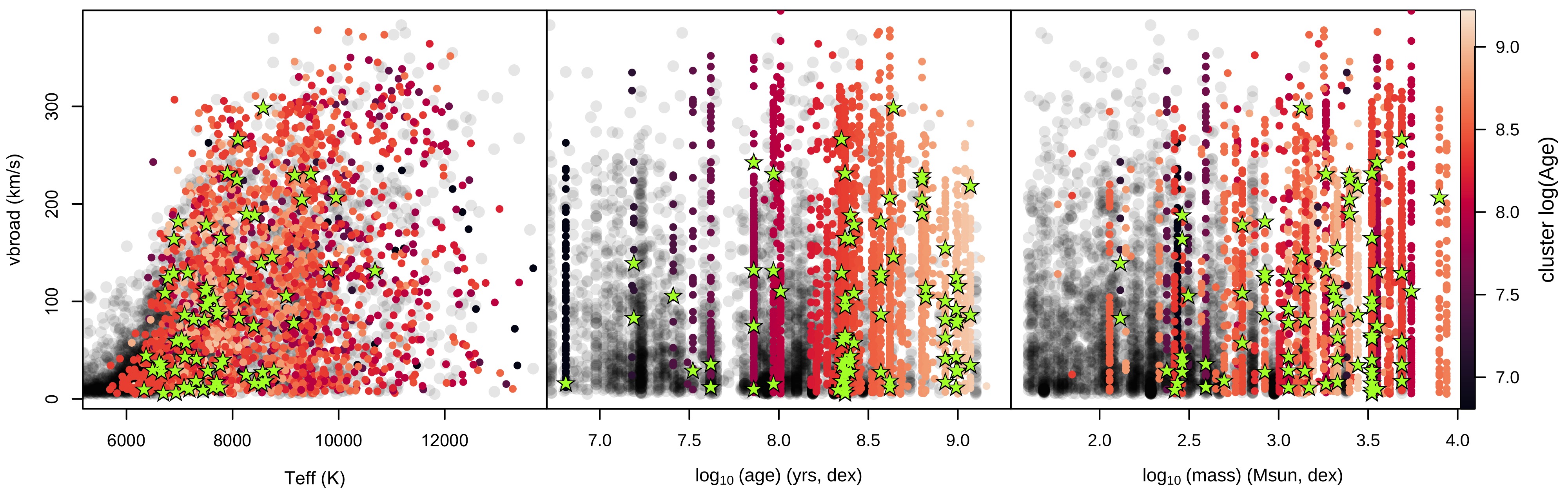}
    \vspace{-0.5 cm}
    \caption{Comparison of the spread in {\tt vbroad} between normal MS stars in clusters showing no sign of an eMSTO (grey points) and stars in the eMSTO region in clusters with a confirmed eMSTO (points colored by age), as a function of \teff\ of the stars (left panel), cluster age (middle panel), and cluster mass (right panel). Binary stars in the eMSTO sample are plotted as green stars.}
   \label{fig:emstovb}
\end{figure*}
% --------------------------------------------

The {\tt vbroad} distribution of the MS and binMS samples is shown in Fig.~\ref{fig:histbinDW}, while the {\tt prot} distribution is shown in Fig.~\ref{fig:histbinPR}. The long tail of fast rotators in the MS {\tt vbroad} sample is dominated by early type stars, where we excluded the stars belonging to extended MS turnoffs (TO, eMSTO, see Sect.~\ref{sec:emsto}), while the flat peak in the MS {\tt prot} distribution is mostly caused by cool and low-mass fast rotators. The MS and binMS distributions have different {\tt vbroad} peaks, at about 15 and 20\,km\,s$^{-1}$, respectively. While the MS peak is in the {\tt vbroad} saturation region, and thus compatible with no rotation, the binMS peak is slightly outside of it. We also observe a different shape in the MS and binMS {\tt vbroad} samples, resembling a sort of systematic shift of $\lesssim$\,30--50\,km\,s$^{-1}$ towards higher {\tt vbroad}. A two-sample KS test -- that the distributions are extracted from the same parent distribution -- returns P-values of 0.0001 and 1.09$\cdot$10$^{-10}$ in the case of {\tt vbroad} and {\tt prot}, respectively. Unresolved spectroscopic SB2 binaries could mimic a higher rotation \citep{simonian20}, thus we draw a line in Fig.~\ref{fig:histbinDW} at 46\,km\,s$^{-1}$, which is twice the resolution of {\em Gaia} RVS. Stars with {\tt vbroad} above the line definitely cannot be unresolved SB2. This is slightly pessimistic, given that in reality {\tt vbroad} starts to saturate at 18--20 rather than 23\,km\,s$^{-1}$ (see App.~\ref{app:rot}). Therefore, a more likely explanation is that in the binMS there are more interacting or post-interaction binaries, which have been spun-up. 

We find comparable proportions of binaries in the MS and binMS samples, i.e., about 7 and 6\%, respectively. We did not spot any obvious difference in the types of binaries found in the two samples. If we assume that the two samples suffer from similar selection biases, we find that the ratio between binaries with q$\gtrsim$0.6 and those with q$\lesssim$0.6 is about 0.8 (see App.~\ref{sec:binfrac} for more details). A flat q distribution would have 0.7. Our data thus support q distributions that include a significant proportion of low mass ratios \citep[see, e.g.][]{kahler99}. In any case, because of the reasons described more in details in App.~\ref{sec:binfrac}, we refrain from computing cluster binary fractions using photometric criteria.

%%%%%%%%%%%%%%%%%%%%%%%%%%%%%%%%%%%%%%%%%%%%%%%%%%

\subsection{Extended Main-Sequence Turnoffs}
\label{sec:emsto}

Extended MS turnoffs (eMSTO) were initially identified in clusters in the Magellanic Clouds as fan-shaped or bimodal turnoff regions \citep{bertelli03,mackey07}. After a long debate about whether the spread was caused by age differences or stellar rotation \citep{goudfrooij14,niederhofer15}, the general consensus finally settled on a spread in stellar rotation, supported by direct measurements in Milky Way open clusters \citep[][and references therein]{bastian18,deng24,cordoni24}. However, there are hints that stellar models including rotation do not always fully account for the observed photometric spreads \citep{georgy13,correnti17,kamann23,cordoni24}. Furthermore, there is no established explanation for why stars should display different rotation. A few possibilities have been put forth \citep[e.g.][]{dantona17,bastian20,wang22}, involving non-canonical stellar evolutionary paths, mediated by binary interactions or early stellar and disk evolution. Observationally, the eMSTO phenomenon was found in several clusters, mostly in the range from 100\,Myrs to 2\,Gyrs, with some exceptions. 

We could identify 96 clusters with a clear eMSTO and 268 with some form of unusual TO regions (labelled as {\tt eMSTO:} in Table~\ref{tab:stars}), significantly increasing the number of known clusters with this feature in the literature, which was a few tens \citep{cordoni24}. We manually selected the eMSTO stars when the spread at the TO was visually larger than on the MS below. We also required that the spread was accompanied by some segregation in the {\tt vbroad} and {\tt vsini} on the blue and red sides of the TO region. However, we included about a dozen clusters without {\tt vbroad} or {\tt vsini} information, either because the photometric signature was sufficiently clear or the case was already known in the literature (one example is NGC\,2818). A few clusters with a clear eMSTO are presented in App.~\ref{app:fig}, sorted by decreasing age. We note that in several clusters there are a few fast-rotating stars ($>$150\,km\,s$^{-1}$) on the blue side of the eMSTO (see e.g. NGC\,5822 in App.~\ref{app:fig}). If the eMSTO morphology was solely driven by stellar rotation, we would expect some low-{\tt vbroad} stars on the red side (because of the sin$i$ projection), but not high-{\tt vbroad} stars on the blue side. These stars are still within the eMSTO region, so we find it unlikely that they are BSS or post-transfer binaries. If this is confirmed, we can consider these stars as additional evidence that a more complex explanation must be invoked, involving more profound structural changes in eMSTO stars \citep[][see Fig.~3]{johnston19}.

One unique advantage of our sample is that we can compare a large number of confirmed eMSTO stars with rotation information ($\simeq$6\,000) with the MS sample, selected to be in the same range of \teff\ and log\,$g$ ($\simeq$8\,000 stars with rotation information), as shown in Fig.~\ref{fig:emstovb}. By comparing Figs.~\ref{fig:histbinDW} and \ref{fig:emstovb}, we observe that the width in the {\tt vbroad} distribution is roughly comparable between the two samples, but the shape of the distribution is different: normal MS stars are more peaked towards low {\tt vbroad} than the eMSTO stars. A K-S test returns a p-value of $<$\,2.2$\cdot$10$^{-16}$. If we assume that the stellar rotation angles are randomly distributed, this suggests a larger proportion of fast rotators in eMSTO stars than in the MS of normal clusters, but we should keep in mind that the {\tt vbroad} distribution within individual clusters can vary \citep[see][]{santos25}.

By inspecting Fig.~\ref{fig:emstovb} we note a few more differences. As a function of T$_{\rm{eff}}$, the MS and eMSTO distributions are roughly similar, in the sense that the upper envelopes are compatible. This shows that, to first order, the maximum {\tt vbroad} is driven by the rotational properties of stars of different masses. However, if we plot {\tt vbroad} as a function of cluster age, we see that above $\simeq$100\,Myr, the eMSTO stars {\tt vbroad} spread seems to globally decrease with increasing age. At older ages the stellar masses in the TO region are lower and we should expect generally lower {\tt vbroad} values. 
This decreasing spread of {\tt vbroad} at old ages seems to contradict the increase of the photometric spread, which is found both when measured as an apparent age spread \citep{niederhofer15} or a pseudo-color spread \citep{cordoni24}. Detailed simulations would be useful to assess whether these two opposite trends are expected on the basis of gravity darkening or whether other structural changes in the star are required to reproduce the observations. Finally, when we plot {\tt vbroad} as a function of cluster mass, we see that at masses above $\simeq$1000\,M$_{\rm{\odot}}$ there is almost no cluster without an eMSTO (grey dots in Fig.~\ref{fig:emstovb}), thus cluster mass seems to play an important role in the emergence of the eMSTO phenomenon and indeed eMSTOs were first discovered in massive clusters in the Magellanic Clouds. We also see very few clusters with an eMSTO below 1000\,M$_{\rm{\odot}}$, but this could in part be due to the fact that eMSTOs are more difficult to detect when there are few stars. 

To test the idea that binary interactions may cause eMSTO \citep{dantona15}, we study 208 binaries (green star symbols in Fig.~\ref{fig:emstovb}) and 131 suspected binaries in the eMSTO sample. Most are eclipsing and SB1, with a couple of G\,Cas variables, and a triple system in NGC\,2516 \citep{mermilliod08,mermilliod09}. We note that they are mostly cooler than 10\,000\,K and mostly found in more massive clusters. The median {\tt vbroad} of the binaries in the eMSTO sample is comparable to that of the whole sample: 90 versus 110\,km\,s$^{-1}$. We do not observe any obvious  preference for the binaries to lie on the red or blue side of the eMSTO, nor any obvious difference in the properties of binaries across the eMSTO, so the present data are inconclusive. We will have to wait for {\em Gaia} DR4 and LSST-Rubin data to be able to further test this idea. 
 
%%%%%%%%%%%%%%%%%%%%%%%%%%%%%%%%%%%%%%%%%%%%%%%%%%

\subsection{Split Main Sequences}
\label{sec:bluems}

% --------------------------------------------
\begin{figure}
    \centering
	\includegraphics[width=\columnwidth]{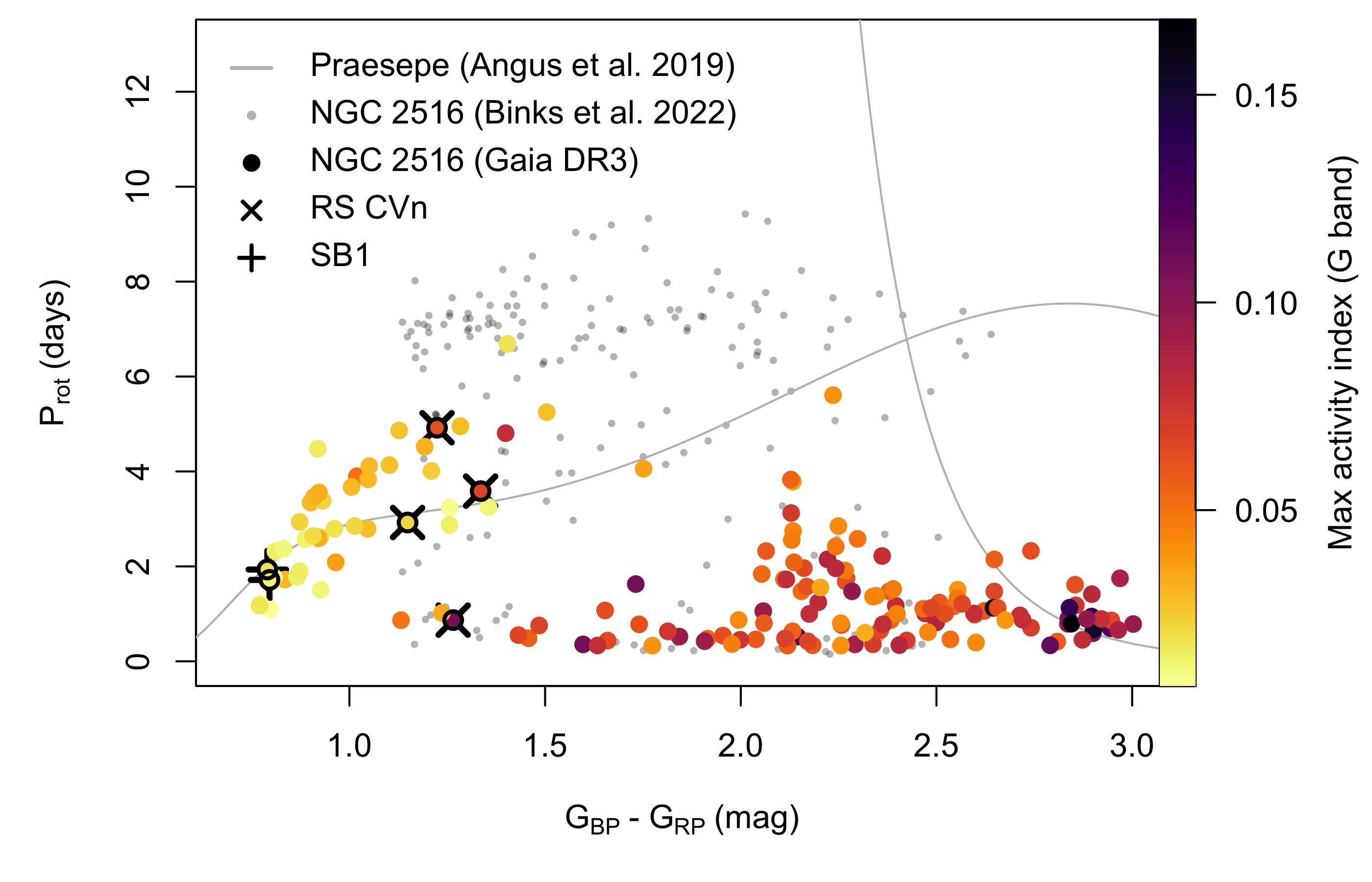}
    \vspace{-0.5 cm}
    \caption{The lower MS of NGC\,2516. Rotation periods from \citet[][grey points]{binks22} and our work (colored points) are compared to the Praesepe period-age relation by \citet[][grey lines]{angus19}. The position of confirmed binaries in our catalogue is marked by crosses. The activity index, A$_{\rm{max}}$, is described in details in App.~\ref{sec:prot}.}
   \label{fig:2516prot}
\end{figure}
% --------------------------------------------

Split MS were first discovered in clusters belonging to the Magellanic Clouds \citep{milone13} and only later found in the Milky Way \citep{cordoni18}. Different explanations were explored, but it is now clear that stars belonging to the two split sequences have different rotation \citep{dantona15,dupree17,sun19}, with the slow rotators on the blue sequence and the fast ones on the red sequence. We identified split MS visually in the CMD of clusters in our sample and found 6 clusters with a clearly splitMS (namely, NGC\,2287, NGC\,2301, NGC\,2422,  NGC\,2516, NGC\,3766, NGC\,6716), to this, we add NGC\,3532 which we classify as eMSTO, but in the literature has also been classified as a splitMS plus an eMSTO \citep{cordoni24}. To our knowledge, only NGC\,2287 and 2422 were previously reported in the literature, besides NGC\,3532. We also found 16 clusters with more uncertain hints of double MS, none of which was previously reported, to our knowledge. We present in App.~\ref{app:fig} three examples, sorted by decreasing age. 

We suggest, judging from clusters with a good sample of {\tt vbroad} measurements, that the split MS is the natural continuation of an eMSTO in lower mass clusters. Split MS in our sample appear always above the transition from radiative to convective cores, i.e. the Kraft break \citep{kraft67}. The age range of clusters with a split MS is contained into the one of eMSTOs. We could thus say that split MS and eMSTO are just two manifestations of the same physical phenomenon \citep[see also][]{sun19}. The only exception could be NGC\,2422, which appears to have a clear split MS but a very narrow eMSTO (if any). But this is also the youngest cluster in our sample, so it is expected to have a narrower eMSTO, given the relation between eMSTO photometric spread and age (see previous section). We note that the masses of clusters hosting split MS are also mostly higher than 1000\,M$_{\rm{\odot}}$, suggesting that cluster mass is an important driving factor, similarly to the case of eMSTO. Finally, we explored the confirmed binary stars in the mentioned clusters: there were very few in the split MS region (from 1 to 4, depending on the cluster). We did not notice any preference for binaries to lie on one or the other sequence nor any preference for a specific type of binary or binary periods \citep[see also][]{kamann21}. 

%%%%%%%%%%%%%%%%%%%%%%%%%%%%%%%%%%%%%%%%%%%%%%%%%%

\subsection{Lower main sequence}
\label{sec:lowMS}

% --------------------------------------------
\begin{figure}
    \centering
	\includegraphics[width=\columnwidth]{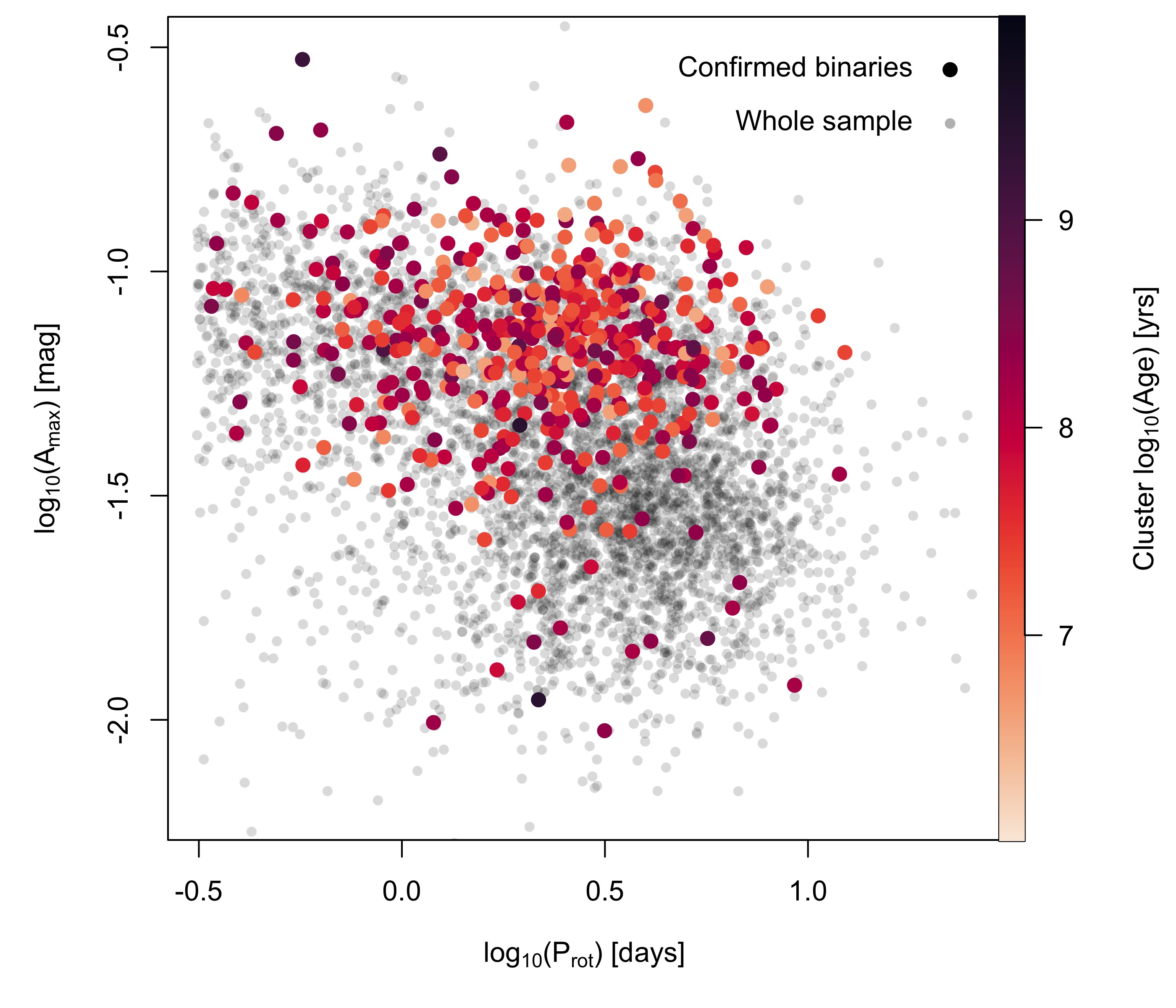}
    \vspace{-0.5 cm}
    \caption{The period-activity diagram for the whole sample of stars in the lower MS (grey points) and the confirmed binaries (colored points).}
   \label{fig:aprot}
\end{figure}
% --------------------------------------------

Rotation periods in {\em Gaia} DR2 and DR3 are mostly measured for stars on the lower MS, where the confusion with other types of variable stars is minimised. Stellar rotation changes with cluster age, as low-mass stars slow down because of magnetic braking \citep[see][and references therein]{angus19}. However, young clusters with bimodal distributions of rotational periods were found, especially among T-Tauri stars, for example in the ONC cluster \citep{attridge92}, NGC\,2264 \citep{lamm05}, IC\,348, and NGC\,2362 \citep{landin21}, but also at slightly older ages \citep{bouvier14}. While slow rotators are well described by gyrochronology models, fast rotators are not. A possible explanation is related to the effect of binary companions or O and B stars radiation on the disks. These reduce or destroy the disk, which may become less able to lock with the star and to slow it down \citep{bouvier93,cuello25}.

A very interesting case is NGC\,2516, which shows a clear bifurcation (or at the very least a spread) between slow and fast rotators in the pre-MS stars (Fig.~\ref{fig:2516prot}). We also notice a possible bifurcation at higher masses in the MS, i.e., bluer than G$_{\rm{BP}}$--G$_{\rm{RP}}$\,=\,1.5\,mag, in the region immediately below the structural transition between stars with convective and radiative cores on the MS ($\simeq$1.1--1.3\,M$_{\odot}$). NGC\,2516 is one of the few clusters in which the whole MS is mapped with both rotation periods and {\tt vbroad} and it also shows a distinct eMSTO as well as a split MS above the transition region and immediately below the eMSTO (see also Sect.~\ref{sec:emsto} and \ref{sec:bluems}). As such, it contains in itself examples of all the bimodalities and spreads discussed in the literature and above. This continuity of properties across the entire range of masses suggests a unifying explanation. One possibility, mentioned above, lies in the role of star-disk interactions.

In an attempt to test the idea that binaries are the underlying cause of the rotation bimodality, for example by disturbing the protostellar disks, we plotted in Fig.~\ref{fig:aprot} the eight confirmed binaries in NGC\,2516, which have both a rotational period and an activity index measurement. The six redder stars are all classified by the {\em Gaia} classifier as RS CVn types, while the two bluer ones are SB1 stars with 0.5 and 8.1\,days of orbital period, i.e., different from their rotational periods, which are around 1.7 and 1.9 days. We note that five of the binaries seem to lie on the fast rotating sequence. However, if we try to generalize this finding using all the clusters in our sample, we encounter some difficulty. Certainly, in clusters above $\simeq$300\,Myr there is no particular tendency of binaries to be faster rotators.

% --------------------------------------------
\begin{figure}
    \centering
    \includegraphics[width=\columnwidth]{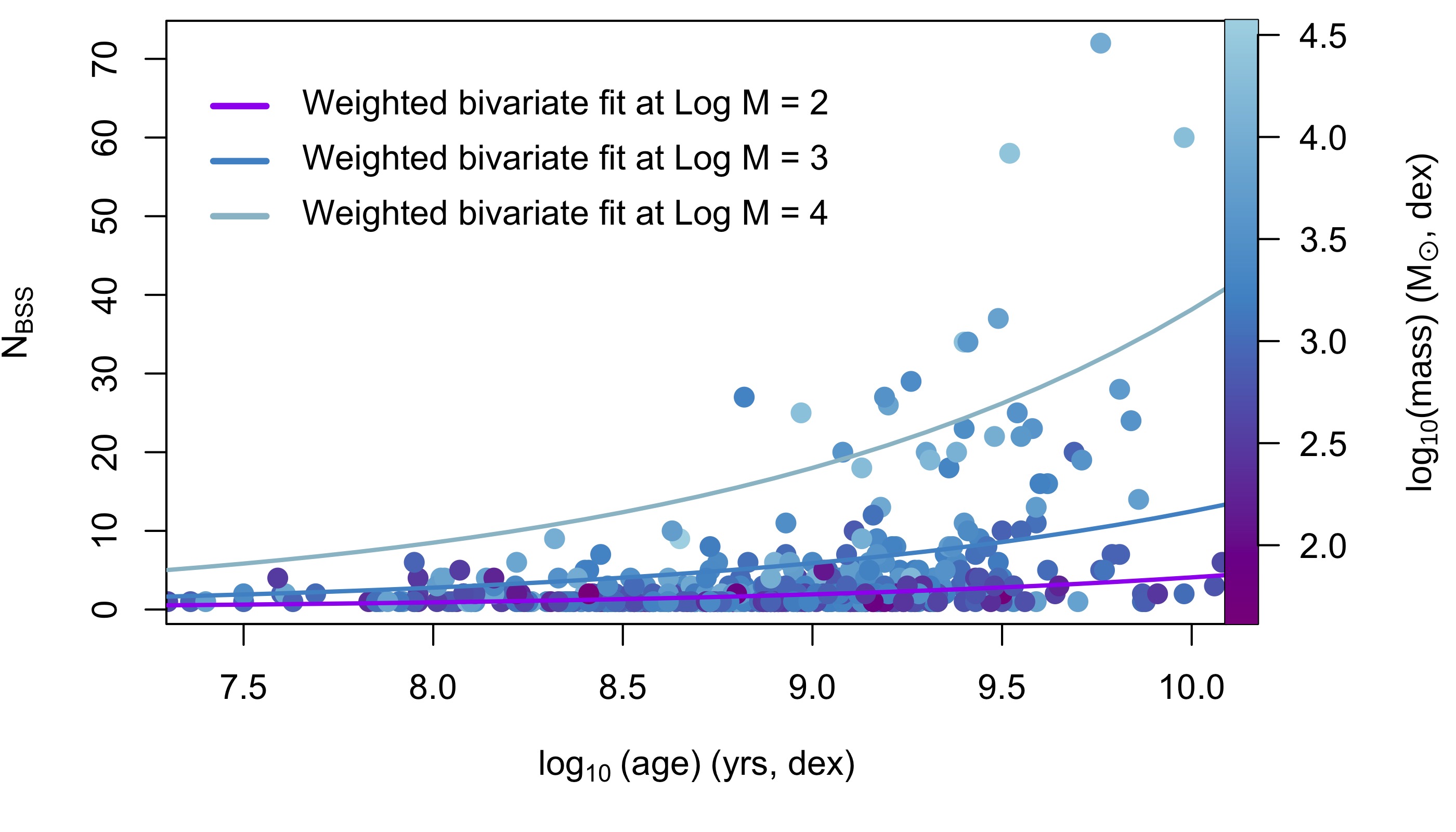}
    \includegraphics[width=\columnwidth]{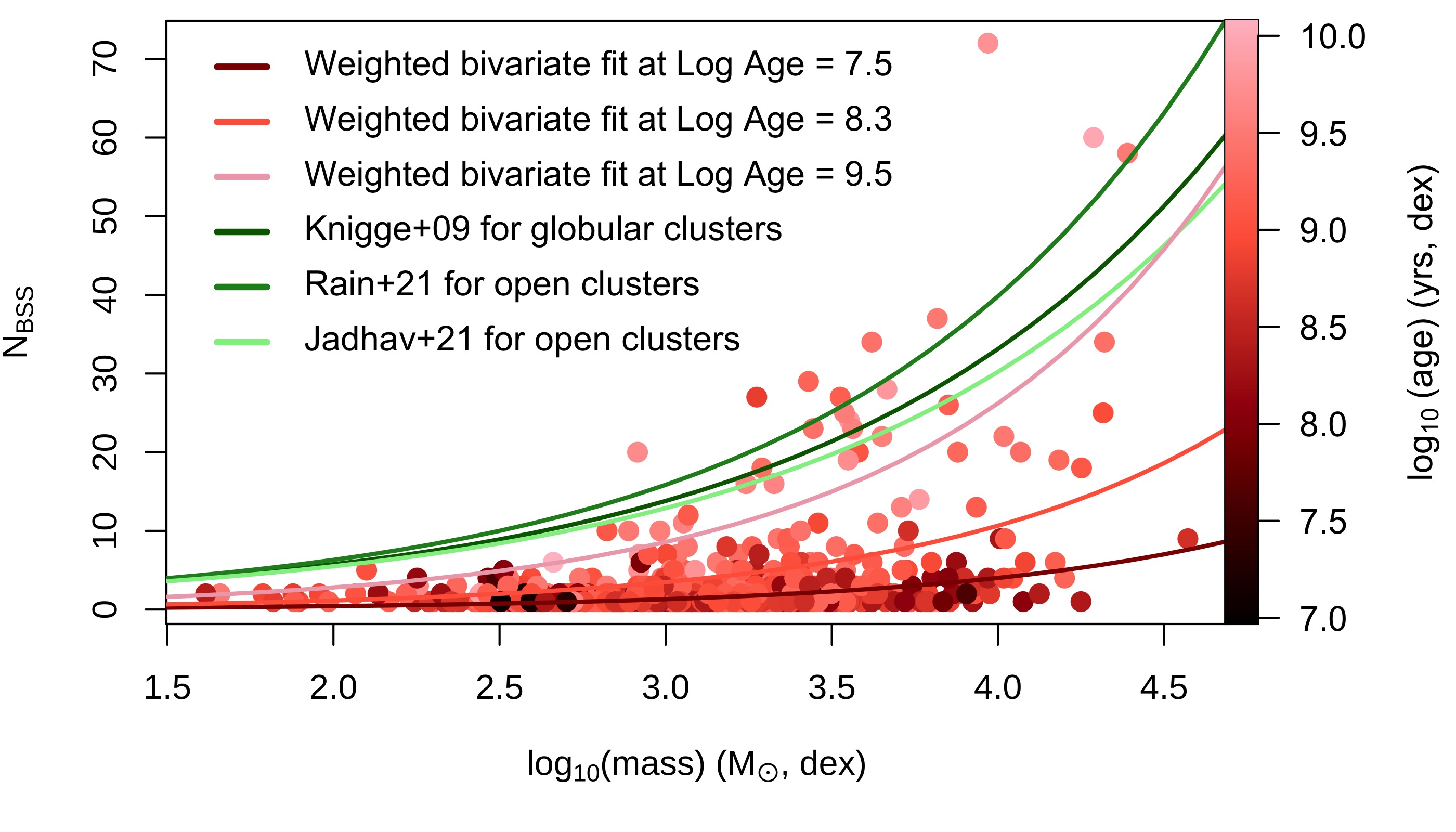}
    \caption{The total number of BSS per cluster as a function of cluster age (top panel) and mass (bottom panel). In the two panels, data points are color-coded by cluster mass and age, respectively. Previous literature efforts of modelling the behaviour as a function of mass by \citet{knigge09}, \citet{jadhav21}, and \citet{rain21}, are plotted in shades of green, as annotated. Our own bivariate fit is plotted in the top panel at fixed Log\,$M$ = 2, 3, and 4 and in the bottom panel at fixed log\,$A$ = 7, 8.3, and 9.5, in the same colorscale as the data points.}
   \label{fig:bss}
\end{figure}
% --------------------------------------------

We show in Fig.~\ref{fig:aprot} the period-activity diagram for our sample of members with {\tt prot}. Thanks to {\em Gaia} DR2, it was possible to identify a population of low-activity fast rotators \citep{lanzafame18,lanzafame19},  which were only found in very small numbers previously \citep{messina08,iwanek19}. It was suggested that this region is populated by binaries, but \citet{lanzafame19} proposed that it is a particular phase of the normal evolution of fast-rotating stars. Few cluster members occupy this region, which contains only two binaries, one RS CVn and one astrometric binary with a period of 616\,days. The overall {\tt prot} distribution of binary stars appears very similar to that of the entire sample. On the contrary, binaries tend to cluster in the region of higher activity, unlike the rest of the stars \citep[see also][]{lanzafame18,lanzafame19}. With our sample we can add that binaries belonging to old clusters tend to be more spread out, especially in activity index, while binaries in younger clusters tend to be more concentrated towards high activity. The binaries in this sample are mostly spectroscopic (SB1 and SB2) and eclipsing. There are also several BY Draconis and RS CVn variables.

%%%%%%%%%%%%%%%%%%%%%%%%%%%%%%%%%%%%%%%%%%%%%%%%%%

\subsection{Blue stragglers}
\label{sec:bss}

Blue Straggler Stars (BSS) were discovered in M\,3 by \citet{sandage53} as stars bluer and brighter than the TO \citep[see also][]{bailyn95}. Since then, they have been found in most star clusters and in the field. They are thought to be the result of binary interactions such as mass transfer, stellar mergers, and stellar collisions \citep{mccrea64,hills76}. Such processes can spin up stars for a certain amount of time, and indeed some BSS have been found to rotate fast in open and globular clusters \citep[e.g.][]{lovisi10,leiner18}. A study based on {\em Gaia} DR2 and astrometric quality indicators sensitive to binarity suggested that BSS have the highest fractions of unresolved binaries in the CMD, compatible with up to 100\% \citep{belokurov20}. Indeed, several open clusters were carefully searched for RV variability in the past \citep[see e.g.][]{mathieu09,gosnell15,geller15}, finding about 75--80\% of binaries in the BSS population, with periods of the order of 1\,000 days. A few eclipsing binaries with shorter periods ($\lesssim$100~days) were also found among candidate BSS \citep{arentoft07,kaluzny14,luo15,Joshi20,rao22}. Compact companions to BSS were found by several authors \citep[e.g.][]{gosnell14,gosnell15,subramaniam16,sindhu19,jadhav21bin,rao22}, further supporting the idea that binary interactions are an important formation channel, which is dominant in open clusters \citep{mathieu25}.

We relied on the BSS catalogs by \citet{jadhav21}, \citet{rain21}, \citet{li23}\footnote{We could not use the new BSS identified by \citet{cui25} because they do not provide the full list.}, and \citet{qin26}, but we manually identified 449 additional BSS or candidate BSS (129 with rotation information) with no literature counterpart. Our final sample contains 1932 BSS (361 with rotation information). As shown in Fig.~\ref{fig:bss}, the total number of BSS per cluster increases with age \citep[as previously noted by][]{ahumada07,rain21,leiner21,li23,cui25}. Our youngest clusters with BSS are about 30~Myr old, i.e. younger than the 100\,Myr reported in the cited studies\footnote{\citet{negueruela23} found that blue supergiants in young clusters are more massive than predicted from single star evolution, and thus there could be BSS in even younger clusters.}. We also observe that the number of BSS increases with cluster mass, as noted previously for both open and globular clusters \citep{knigge09,jadhav21,rain21}. With our simultaneous mass and age determinations, we note that both need to be high, i.e. M\,$\gtrsim$\,1000\,M$_{\rm{\odot}}$ and age\,$\gtrsim$\,500\,Myr respectively, to have more than $\simeq$5--10 BSS in a given cluster. We computed a weighted bivariate fit that appears to capture  more closely the observed behaviour, compared to previous univariate fits \citep{rain21,jadhav21}
$$N_{\rm{BSS}} = 10^{~-3.62(\pm0.26)~+~ 0.48(\pm0.03)~{\rm Log} M ~+~ 0.33(\pm0.03)~{\rm Log}A}$$
\noindent where $M$ and $A$ are mass and age, respectively and $N_{\rm{BSS}}$ is the number of BSS in each cluster. The fits are reported in Fig.~\ref{fig:bss}.

In Fig.~\ref{fig:histbinDW} we examine the overall {\tt vbroad} distribution of 77 BSS. We observe that among the studied populations, BSS are the ones which differ the most from the MS sample. Their distribution has an irregular shape, which could be an indication of an underlying multimodality or just the result of the relatively smaller sample. We suspect that the main peak at $\simeq$50\, km\,s$^{-1}$ could be at least in part caused by unresolved SB2, but the excess of fast rotators at $\simeq$150\,km\,s$^{-1}$ is certainly not made by unresolved binaries. We find a larger median {\tt ruwe}\footnote{The renormalized unit weight error in the {\em Gaia} catalog indicates extra noise in the astrometric solution for a given source, that can also be caused by binary companion, among other reasons.} of 1.14 in the BSS rotating slower than 100\,km\,s$^{-1}$ than in the faster ones, where we find only 0.98. This could be taken as supporting evidence that the $\simeq$50\, km\,s$^{-1}$ peak is dominated by astrometric binaries, but needs confirmation. To compare with globular clusters, we note that \citet{ferraro23} found 40\% of the BSS rotating faster than $>$40\,km\,s$^{-1}$ in loose clusters, while only 4\% in high-density globular clusters. Our fraction of BSS in open clusters rotating faster than 40\,km\,s$^{-1}$ is 76\%. This makes sense if we consider that all open clusters are certainly loose compared to globular clusters. In the case of globular clusters, this was interpreted as evidence that BSS are still forming in loose clusters. This could be the case also in open clusters, i.e. fast rotating BSS would be recently formed, but with the caveat that we also have some more massive BSS in our sample, which could spin down more slowly than low mass ones \citep[see][for FGK type BSS]{leiner18} and thus should not be interpreted as more recently formed. 

Finally, we report that 97 BSS are found to be binaries or suspected binaries in our sample (i.e. about 6\%). They are mostly classified as eclipsing binaries, but there are also some SB1 and SB2, and some might have hot companions, judging from their UV excess \citep{jackim24}. A few are classified in {\em Gaia} with acceleration solutions, which means that the period must be comparable to or longer than the 34 months covered in DR3. The few having period determinations are always below 4\,days, only one has 233\,days. Four stars are contact binaries and one may be in a common envelope phase. These short period binaries are not common among BSS, which is compatible with the small fraction found here. They may be indicative of case A mass transfer\footnote{Case A mass transfer happens when the donor is on the MS, while case B on the RGB, and case C on the asymptotic giant branch.}, with low-mass MS companions expected \citep{rao22,sandquist03}. Longer period BSS binaries, about 75-80\% in open clusters in the literature \citep{mathieu25}, have often WD-like companions, and would more likely be the result of case B or C mass transfer, but for the moment they are out of reach for {\em Gaia}.

%%%%%%%%%%%%%%%%%%%%%%%%%%%%%%%%%%%%%%%%%%%%%%%%%%

% --------------------------------------------
\begin{figure}
    \centering
	\includegraphics[width=\columnwidth]{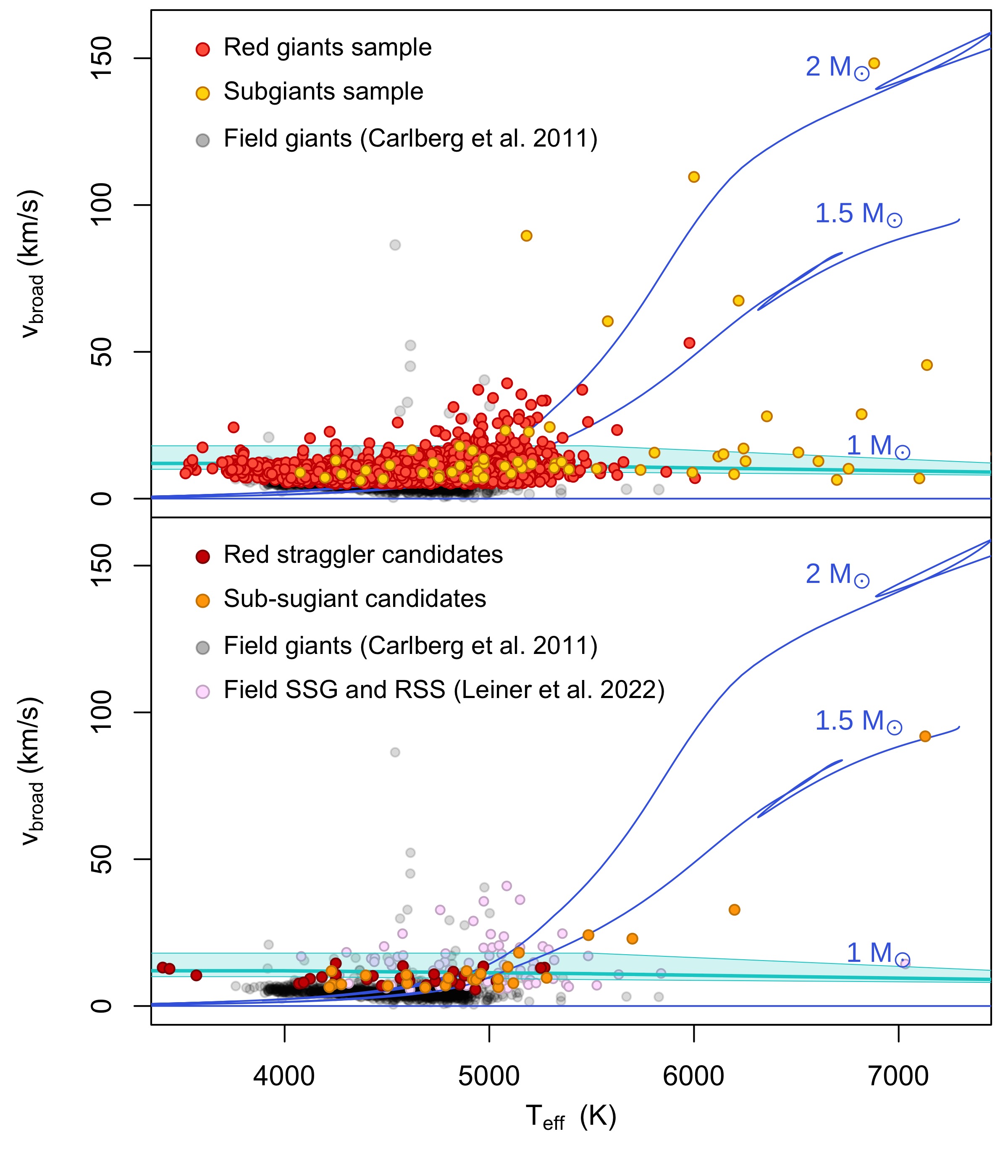}\\
	\vspace{-0.3cm}
    \caption{{\em Top panel.} The RGB (red points) and SGB (yellow points) samples are compared with the sample of field giants by \citet{carlberg11} in the \teff\ {\tt vbroad} plane. {\em Bottom panel.} Same as the top panel, but for the RSS (dark red points) and SSG (orange points) candidates. Here we also plot the {\em Gaia} measurements for the sample of field SSG and RSS (thistle points) by \citet{leiner22}. In both panels, the cyan line and shaded area represent the region where {\em Gaia} {\tt vbroad} starts to be overestimated, with its uncertainty range \citep[][Table~2]{fremat22}. The MIST rotational models \citep{choi16} with $\Omega_{\rm{ZAMS}}/\Omega_{\rm{crit}}$\,=\,0.4 are also plotted in blue for reference.}
   \label{fig:giants}
\end{figure}
% --------------------------------------------
% --------------------------------------------
\begin{figure}
    \centering
    	\includegraphics[width=\columnwidth]{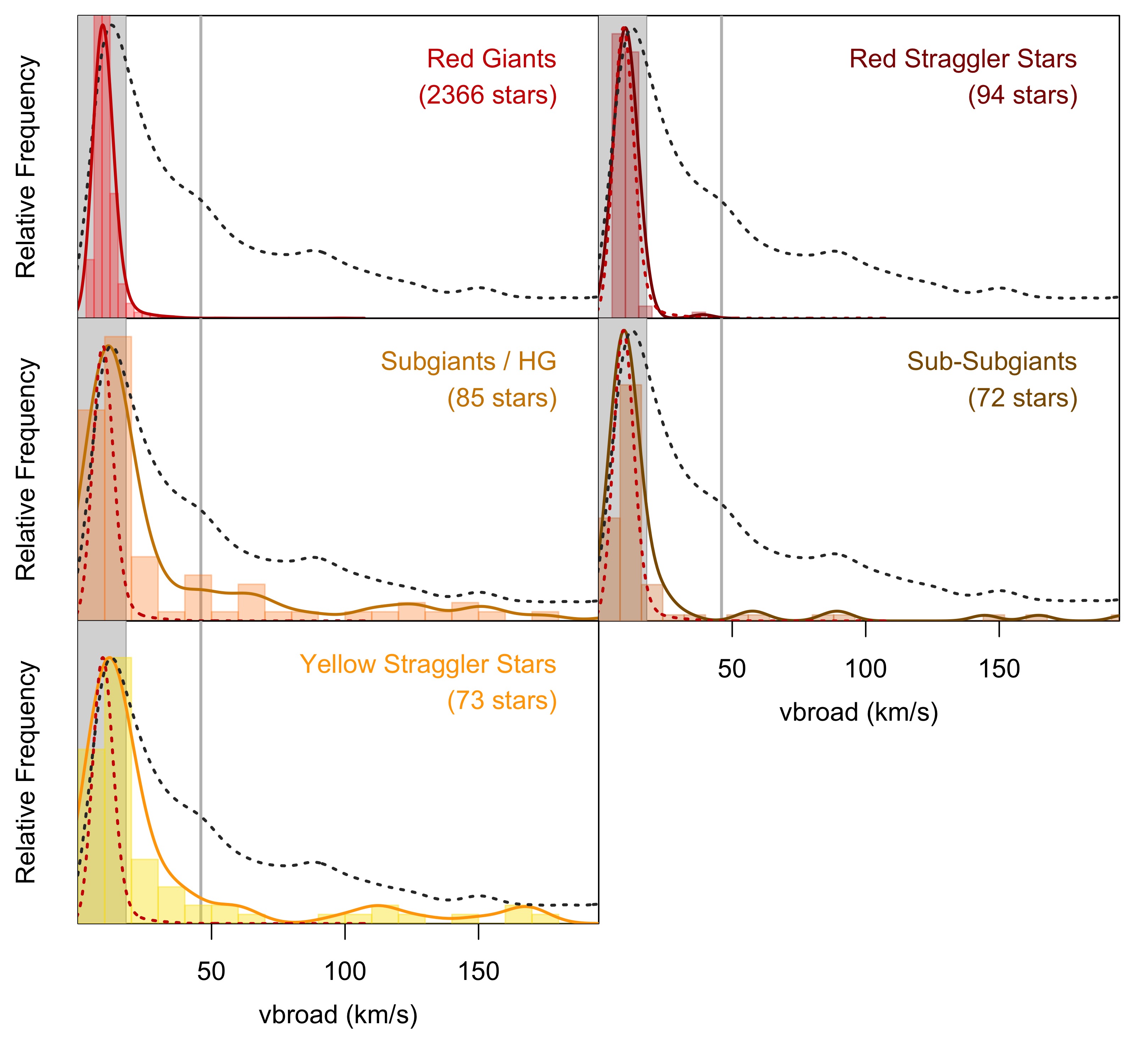}
    \vspace{-0.3cm}
    \caption{Similar to Fig.~\ref{fig:histbinDW} but for the samples of giants, as annotated.
    }
   \label{fig:histbinG}
\end{figure}
% --------------------------------------------

\subsection{Red giants and subgiants}
\label{sec:giants}

Fast-rotating  ($>$\,10\,km\,s$^{-1}$) red giant branch stars (RGB) are rare and are thought to be byproducts of binary interactions \citep{carlberg11,gaulme20,patton24}, because the expanded envelopes of giant stars lead to low projected rotational speed, generally $<$\,5\,km\,s$^{-1}$. Among the interactions with (sub)stellar companions that can transfer angular momentum there are tidal synchronization, mass transfer, and mergers. Thus, in some cases we may be able to observe a companion, while in the case of mergers the star will be single \citep[see, e.g.][and references therein]{carlberg11}. 

We selected a sample of rotationally characterised red giants and subgiant branch stars (SGB) by visually inspecting the CMD of each cluster having at least one {\tt vbroad} or {\tt vsini} measurement. We used BaSTI isochrones \citep{hidalgo18} to guide the eye in the case of less populated clusters, but the candidates were selected only when the RGB and SGB sequences were clearly visible in the CMD and we did not use the isochrones in the actual selection \citep[see, e.g.,][]{ramos24}. Our final sample of rotationally characterised red giants and subgiants contains 2369 and 85 stars, respectively, which are displayed in the top panel of Fig.~\ref{fig:giants}. Even considering the limitations of the {\em Gaia} {\tt vbroad} measurements (cyan region in Fig.~\ref{fig:giants}), our sample overlaps the one of field giants by \citet{carlberg11}. The difference in spectral resolution between the two samples can be appreciated, because the \citet{carlberg11} sample reaches down to a broadening of almost 0\,km\,s$^{-1}$ while the {\em Gaia} one barely reaches 5\,km\,s$^{-1}$. We also observe that the RGB distribution is different from the dwarfs samples in Figs.~\ref{fig:histbinDW} and \ref{fig:histbinPR}. The {\tt vbroad} distribution of RGB and SGB stars is shown in Fig.~\ref{fig:histbinG} and is very strongly peaked at 9.6\,$\pm$\,2.6~km\,s$^{-1}$. This is determined by the RVS resolution limits \citep{fremat22} and does not represent the intrinsic rotation of the stars, thus we focused our analysis on the long tail of fast-rotating stars, reaching 98.3~km\,s$^{-1}$. 

We selected fast rotating giants as those being more than 3\,$\sigma$ faster than the average, i.e. {\tt vbroad}\,$>$\,17.4. We obtained a sample of 79 stars or 4\%, which is comparable with previous findings in the literature \citep{carlberg11}. Of these, 26 (33\%) are flagged as variable or suspect variable stars (see Sect.~\ref{sec:var}) and 9 (11\%) as binary or suspect binaries (see Sect.~\ref{sec:bin}). For reference, the whole giants sample has a variable star fraction of 14\% and binary fraction of 11\%. Most classifications are uncertain or generic (long period, rotational). Notably, there are three fast rotating giants classified by {\em Gaia} as RS CVn, but not parameterised; five are classified by {\em Gaia} or spectroscopic surveys as SB1; and two are classified as eclipsing binaries. The available periods range from 8 to $>$\,10$^4$\,d, with a median of about 100\,d. Because our binary characterization may be biased and incomplete, we also tried another approach. An analysis by \citet{patton24} employed indirect criteria such as velocity spreads of APOGEE data and {\tt ruwe} in {\em Gaia} to estimate the number of binaries among fast-rotating giants in APOGEE, finding striking fractions of 75\% or higher. Our giants sample has a median {\tt ruwe}\,=\,1.01\,$\pm$\,0.12. Therefore, an appropriate threshold would be 1.37, different from the value of 1.2 used by \citet{patton24}. In any case, using the two thresholds, we find 6\% and 11\% of suspect binaries, respectively, which is compatible with the fraction computed over the entire giants sample, 5\% and 12\%, respectively. 

We note that the saturation threshold of {\em Gaia} DR3 {\tt vbroad} prevents us from separating the intermediate rotators \citep[or anomalous rotators, with 5--10\,km\,s$^{-1}$,][]{tayar15} from the slow rotating giants. This could be one of the causes of the tension between our results on the binary fractions and those by, e.g. \citet{patton24}. Thus, a possible interpretation is that our sample of fast rotators is dominated by mergers and that binaries may dominate in the intermediate sample that we cannot discern with our data. Another possibility is that, considering that mergers should be more common in the cluster environment than in the field, our cluster sample of giants is intrinsically different from the field samples in the literature, containing relatively more mergers and therefore less binaries.

%%%%%%%%%%%%%%%%%%%%%%%%%%%%%%%%%%%%%%%%%%%%%%%%%%

\subsection{Red stragglers and sub-subgiants}
\label{sec:ssg}

Red stragglers (RSS) and Sub-subgiants (SSG) occupy a region of the CMD which is respectively redder than the red giant branch or fainter than the sub-giant branch. The first two SSG were discovered in M\,67 by \citet{mathieu03}. Several years later a census of the 65 known SSG and RSS in open and globular clusters was published by \citet{geller17}, and a sample of 448 RSS and SSG in the field population was presented by \citet{leiner22}. Several of these stars are X-ray emitters, binaries, variables, spotted, and/or display H$_{\rm{\alpha}}$ emission. Moreover, RS CVn variables are found in the known samples \citep{dixon25}. According to \citet{leiner22}, {\tt prot} peaks at about 5\,days in the field sample, with a long tail reaching tens of days. The distribution of {\tt prot} for field and cluster SSG appear compatible with each other. Two different explanations were put forward in the above cited studies for their non-canonical position in the CMD. The favoured one is that SSG and RSS are spun-up by binary interactions and develop magnetic fields, appearing fainter and redder because of spots and possibly radius inflation. The second is that they are undergoing important mass transfer episodes which bring them temporarily out of equilibrium. 

We selected 158 (rotationally characterised) candidate SSG and RSS from individual clusters CMDs. When the RGB and SGB were sufficiently populated, we selected our candidates as 3\,$\sigma$ outliers. When the RGB and SGB were scarcely populated, we relied on BaSTI isochrones \citep{hidalgo18}, and thus on cluster parameters, to guide our selections. Unfortunately, we have no candidates in common with \citet{geller17}, because their sources are too faint to be rotationally characterised in {\em Gaia} DR3. The sample is presented in Fig.~\ref{fig:giants} (bottom panel), where it is compared with the distribution of field giants by \citet{carlberg11} and the {\em Gaia} parameters of field SSG and RSS by \citet{leiner22}. The fraction of fast rotating stars in our sample agrees well with the field sample by \citet{leiner22}. If we cross-match the field sample with {\em Gaia}, we find 5\% of the stars with {\tt vbroad}\,$>$\,17.4\,km\,s$^{-1}$, while if we do the same with our sample, we find 8\%. Moreover, it was found that {\tt prot} tends to increase from SSG to RSS \citep{leiner22}, compatibly with stellar evolution expectations. Our sample confirms this finding, as can be seen from Fig.~\ref{fig:giants} but also from Fig.~\ref{fig:histbinG}: the RSS {\tt vbroad} distribution is more peaked and has virtually no tail towards higher values, unlike the SSG sample. 

Concerning the possible binary nature of our candidates, we note that the entire sample of RSS and SSG has a fraction of variable stars of 7\% and of binary stars of 17\%. These figures are higher than, but comparable to those of the combined RGB and SGB sample, which are 14\% and 11\%. The types of binaries found among RSS and SSG candidates are typically SB1 spectroscopic binaries. The only fast rotating RSS is reported by {\em Gaia} DR3 as an RS CVn type, but is not parametrised. Previous reports by the ATLAS survey \citep{heinze18} also classify it as a possible eclipsing binary with a sinusoidal light curve. The {\tt ruwe} of these RSS and SSG candidates is indistinguishable from that of normal giants. If {\tt ruwe}\,$>$\,1.37 was taken as an indirect indication of binarity, it would suggest a binary fraction of 14\%. In conclusion, with the data in hand, we can only report a slight increase (17\% versus 11\%) in the binary discovery fraction of RSS and SSG with respect to the normal red giants.

%%%%%%%%%%%%%%%%%%%%%%%%%%%%%%%%%%%%%%%%%%%%%%%%%%

\subsection{Yellow straggler candidates}
\label{sec:yss}

Evolved BSS should appear brighter than the sub-giant branch and redder than BSS. In the literature they are thus called either red or yellow straggler stars \citep{sales14}. To avoid any confusion with the RSS (Sect.~\ref{sec:ssg}), we adopt here the term yellow straggler star (YSS). Unfortunately, the region where evolved BSS are expected to lie also hosts evolved binaries of various kinds, and the continuation of the equal-mass binary sequence (Sect.~\ref{sec:binms}). Indeed, several of the known YSS have shown A-type companions, identified by the presence of Balmer lines and veiling in the blue spectral regions \citep{sales14,dasilveira18,martinez20,rani23}. In these cases, it is believed that the yellow color is driven by the contamination of the blue companion, making YSS bluer than RGB stars, but still redder than the typical BSS. For all these reasons, we will refer to our YSS sample as YSS candidates in the following.

We relied on the YSS selected by \citet{rain21} and \citet{li23}. We further selected new YSS candidates as stars redder than the TO, bluer than the RGB, and brighter than the SGB. We only used clusters with sufficiently populated sequences and we did not use isochrones to compensate for the lack of data. Our YSS candidates sample counts 134 stars, doubling the literature samples. In Fig.~\ref{fig:histbinG} we compare their {\tt vbroad} distribution with that of other samples of giants and subgiants, discussed above. The YSS candidates rotating faster than 17.4\,km\,s$^{-1}$ (see Sect.~\ref{sec:giants}) are $\simeq$38\%, compared to 4\% for normal giants. We found 23 potential or confirmed binaries in the YSS sample (17\%). Almost all are classified as SB1 \citep{pourbaix04,gavras23}, a few are classified as eclipsing binaries or RS CVn in {\em Gaia} DR3, but they are not parametrised. One has UV excess in GALEX \citep{jackim24}. When available, periods range from one to more than 1000\,days. Only one of the binaries is also rotating faster than 17.4\,km\,s$^{-1}$. If all the rotating YSS candidates are assumed to be binaries, this would bring the (still uncorrected) binary fraction of YSS candidates to $\simeq$46\%. This result supports the idea that this portion of the CMD is dominated by binaries and their byproducts. Unfortunately, as mentioned, we cannot cleanly separate YSS from other types of binaries which are known to populate this region.

%%%%%%%%%%%%%%%%%%%%%%%%%%%%%%%%%%%%%%%%%%%%%%%%%%

\section{Summary and conclusions}
\label{sec:conc}

In this work, we have selected and characterised a sample of more than 700\,000 members of almost 6\,000 open clusters. We have employed {\em Gaia} DR3 information of stellar rotation, both from the width of RVS spectra and from the lightcurves of spotted rotating stars, to study the behaviour of various stellar samples across the CMD. 

Our main results can be summarised as follows:
\begin{itemize}
\item{We have classified a large number of stars in the MS and binMS and we found a statistically significant excess of rotation in the binMS sample. The binMS sample contains a similar proportion of binaries as the the MS ($\simeq$6\% compared to 7\%). To clean the samples from abnormally fast rotating stars, we produce a sample of about 100 new candidate BL, which we make available for further study.}
\item{We have significantly increased the number of clusters with an eMSTO or a split MS compared to the previous literature, bringing it to about 100, plus a few hundred clusters with suspected complex TO regions. We support the idea that the photometric spread is not entirely explained by the {\tt vbroad} spread, as previously suggested in the literature. We further observe that most clusters with a mass higher than about 1000\,M$_{\rm{\odot}}$ display an eMSTO. We also notice that the spread in {\tt vbroad} seems to decrease with age for older clusters, in apparent contradiction with the photometric spread, which increases with age. This may not be a true inconsistency, but it needs to be explored with the help of models.}
\item{In the lower main sequence, we observe that the confirmed binaries can be slow or fast rotators, but most of them have high activity, as measured from the amplitude of the light curve, as was previously noted in the literature. We further note that binaries in older clusters extend to lower activity levels compared to younger clusters.}
\item{We discovered several hundreds of new stars belonging to various exotic populations. In the case of BSS, we obtained a combined sample of new discoveries and literature detections to almost 2000 BSS. We observe that the clusters with more than 5--10 BSS generally have both high mass ($\gtrsim$1000\,M$_{\rm{\odot}}$) and old age ($\gtrsim$1\,Gyr) and we provide a bivariate fit that reproduces the behaviour better than previous univariate fits. The {\tt vbroad} distribution significantly differs from the one of MS stars and presents an irregular shape. The fraction of stars rotating more than 40\,km\,s$^{-1}$ is about 75\%. We found several binaries with relatively short periods among BSS (about 6\%), compatible with previous literature studies. Longer period binaries are for the moment out of reach of {\em Gaia}.}
\item{Among red giants and subgiants we found a small group of abnormally fast rotators (about 4\%), compatible with previous findings for field stars, while the rest of the sample displays the expected rotational behaviour. These fast rotating stars can only be the result of recent binary interactions and indeed they display higher proportions of known binaries than the slowly rotating giants. }
\item{The samples of YSS candidates, RSS, and SSG display more prominent tails of fast rotating stars, compared to the RGB sample. The percentage of binaries in these populations is slightly higher than in the RGB sample. About 33\% of fast-rotating YSS candidates are binaries.}
\end{itemize}

Our bird-eye exploration of the properties of rotating stars in open clusters revealed a few interesting new facts that are worth following up with dedicated studies. We hope that this new and increased sample will help in solving some of the long-standing open problems in the field. The future is bright. On the one hand, the spectroscopic surveys already contain a treasure trove of {\tt vbroad} measurements that only awaits to be explored in a synergic way, similarly to what has been done in projects such as the Survey of Surveys \citep{tsantaki22,turchi25}. On the other hand, the upcoming {\em Gaia} DR4 data are expected to improve significantly on the quantity and quality of the measurements and especially on the inventory of binary stars, which are expected to be closely linked with stellar rotation. LSST-Rubin will additionally soon provide an overwhelming amount of high-quality light-curves \citep{ivezic19}.

%%%%%%%%%%%%%% ACKNOWLEDGEMENTS %%%%%%%%%%%%%%%%%%

\begin{acknowledgements}

{\bf People.} We would like to thank the following colleagues for useful comments and suggestions: Eleonora Bianchi, Katia Biazzo, Santi Cassisi, Michele Fabrizio, Yves Fr\'emat, Alessandro Lanzafame, Emily Leiner, Bob Mathieu, Fr\'ed\'eric Royer, Claudia Toci.

{\bf Funds.} Co-funded by the European Union (ERC-2022-AdG, {\em "StarDance: the non-canonical evolution of stars in clusters"}, Grant Agreement 101093572, PI: E. Pancino). Views and opinions expressed are however those of the author(s) only and do not necessarily reflect those of the European Union or the European Research Council. Neither the European Union nor the granting authority can be held responsible for them. 
EP acknowledges financial support from PRIN-MIUR-22: CHRONOS: adjusting the clock(s) to unveil the CHRONO-chemo-dynamical Structure of the Galaxy” (PI: S.~Cassisi) funded by the European Union -- Next Generation EU.
VVJ acknowledges support through grant 26-21774S from the Czech Grant Agency.

{\bf Data.} This work has made use of data from the European Space Agency (ESA) mission Gaia (\url{https://www.cosmos.esa.int/gaia}), processed by the Gaia Data Processing and Analysis Consortium (DPAC, \url{https://www.cosmos.esa.int/web/gaia/dpac/consortium}). Funding for the DPAC has been provided by national institutions, in particular the institutions participating in the Gaia Multilateral Agreement. 

Funding for the Sloan Digital Sky Survey IV has been provided by the Alfred P. Sloan Foundation, the U.S. Department of Energy Office of Science, and the Participating Institutions. SDSS-IV acknowledges support and resources from the Center for High Performance Computing at the University of Utah. 

Guoshoujing Telescope (the Large Sky Area Multi-Object Fiber Spectroscopic Telescope LAMOST) is a National Major Scientific Project built by the Chinese Academy of Sciences. Funding for the project has been provided by the National Development and Reform Commission. LAMOST is operated and managed by the National Astronomical Observatories, Chinese Academy of Sciences. 

{\bf Tools.} Most of the plotting and data analysis was carried out using R \citep{R,data.table}. This research has made use of the SIMBAD database \citep{simbad} and the VizieR catalogue access tool \citep{vizier}, both operated at CDS, Strasbourg, France. Preliminary data exploration relied on TopCat \citep{taylor05}. 
\end{acknowledgements}

%%%%%%%%%%%%%%%% BIBLIOGRAPHY %%%%%%%%%%%%%%%%%%%%

\bibliographystyle{aa} % style aa.bst
\bibliography{OCSwDR3} % your references Yourfile.bib

%%%%%%%%%%%%%%%%% APPENDICES %%%%%%%%%%%%%%%%%%%%%

\appendix

%%%%%%%%%%%%%%%%%%%%%%%%%%%%%%%%%%%%%%%%%%%%%%%%%%

\section{More details on sample selection}
\label{app:sample}

To integrate the missing clusters in the \citet{hunt24} catalogue (see Sect.~\ref{sec:ocmaster}), we created our own collection of member lists, using the following large literature studies:

\begin{itemize}
    \item{\citet{cantat20_430k} and \citet{cantat20_250k}, which contain more than 480\,000 stars belonging to 2017 clusters; they also homogeneously re-analised clusters previously published by various authors \citep[e.g.][and others]{sampedro17,cantat18,cantat18b,castro18,castro19}.}
    \item{\citet{sim19} did not provide any members lists, so we could not use their work, but we note that of their 207 clusters, 137 are included in \citet{cantat20_430k} and \citep{cantat20_250k}.}
    \item{\citet{castro22} provided 25466 members for 628 clusters (615 new), using {\em Gaia} EDR3.}
    \item{Among the clusters by \citet{ferreira19,ferreira20,ferreira21}, we found 26 not included above, with 4539 new member stars.}
    \item{\citet{he21} found 74 new clusters with 3754 members.}
    \item{\citet{hao22} provided 704 (378 new) clusters with 19425 member stars, using {\em Gaia} EDR3.}
\end{itemize}

Because some of the above studies were based on {\em Gaia} DR2 \citep{gdr2}, we computed a DR2-DR3 cross-match using the software by \citet{marrese17,marrese19} and used the best matches as our {\em bona fide} identifications. We identified more than 200\,000 member stars in our literature collection that were missing from the \citet{hunt24} list.

As mentioned in Sect.~\ref{sec:starmaster}, in most cases the CMDs obtained with the \citet{hunt24} list appear cleaner and better defined, therefore we did not want to blindly merge the two lists. We thus adopted the following approach: we visually inspected the CMDs and vector-point diagrams of all the relevant clusters to identify the ones, like NGC\,7789, for which we considered the \citet{hunt24} selection to be too restrictive in the coordinates or proper motion-parallax planes, compared to the distribution of stars in the region. Only for these clusters, we added back to our masterlist all the stars that were in our literature collection, but were missing from \citet{hunt24}. These extra stars were later retained only if they had rotational information in {\em Gaia} DR3 \citep[][see also Sect.~\ref{sec:rot}]{gdr3}, after checking that the {\em Gaia} DR3 positions, motions, colors, and magnitudes of the missing stars were still compatible with being cluster members. We also retained all the member stars of the 51 missing clusters in \citet{hunt24} discussed in Sect.~\ref{sec:ocmaster} above, regardless of their rotational characterization. In this way, we added more than 10\,000 stars to our masterlist.

Notably, the list contains some repeated stars, which were assigned as members to different nearby clusters, with different membership probabilities. We decided to keep these duplicates as separate entries in our masterlist. 

%%%%%%%%%%%%%%%%%%%%%%%%%%%%%%%%%%%%%%%%%%%%%%%%%%

\subsection{Gaia quality selections and additional information}
\label{sec:gaiaqc}

Using the {\em Gaia} DR3 {\tt source\_id} for our stars, we extracted relevant data from the following {\rm Gaia} Tables: 
\begin{itemize} 
\item{{\tt gaia\_source},}
\item{{\tt astrophysical\_parameters},}
\item{{\tt vari\_summary},}
\item{{\tt vari\_rotation\_modulation}, }
\item{{\tt nss\_two\_body\_orbit},} 
\item{{\tt nss\_acceleration\_astro}, }
\item{{\tt nss\_non\_linear\_spectro}. }
\end{itemize}

All stars matching the following criteria, based on information in the above Tables, were rejected from our masterslist:
\begin{itemize}
\item{sources missing G, BP, or RP magnitudes}
\item{sources with {\tt in\_qso\_candidates=TRUE}}
\item{sources with {\tt in\_galaxy\_candidates=TRUE}}
\item{sources with {\tt classprob\_dsc\_combmod\_quasar$>$0.5}}
\item{sources with {\tt classprob\_dsc\_combmod\_\-galaxy$>$0.5}}
\end{itemize}

Apart from these very basic selection criteria, which removed a few hundred stars from the masterlist, we preferred to flag stars rather than to filter them out. This is because we focus on rotating stars, which can have altered colors and magnitudes, whose RV determination could be less precise, and which might be variable or binary and have less precise astrometry as well.

%%%%%%%%%%%%%%%%%%%%%%%%%%%%%%%%%%%%%%%%%%%%%%%%%%

\section{Details on rotational characterization}
\label{app:rot}

% --------------------------------------------
\begin{figure}
    \centering
	\includegraphics[width=\columnwidth]{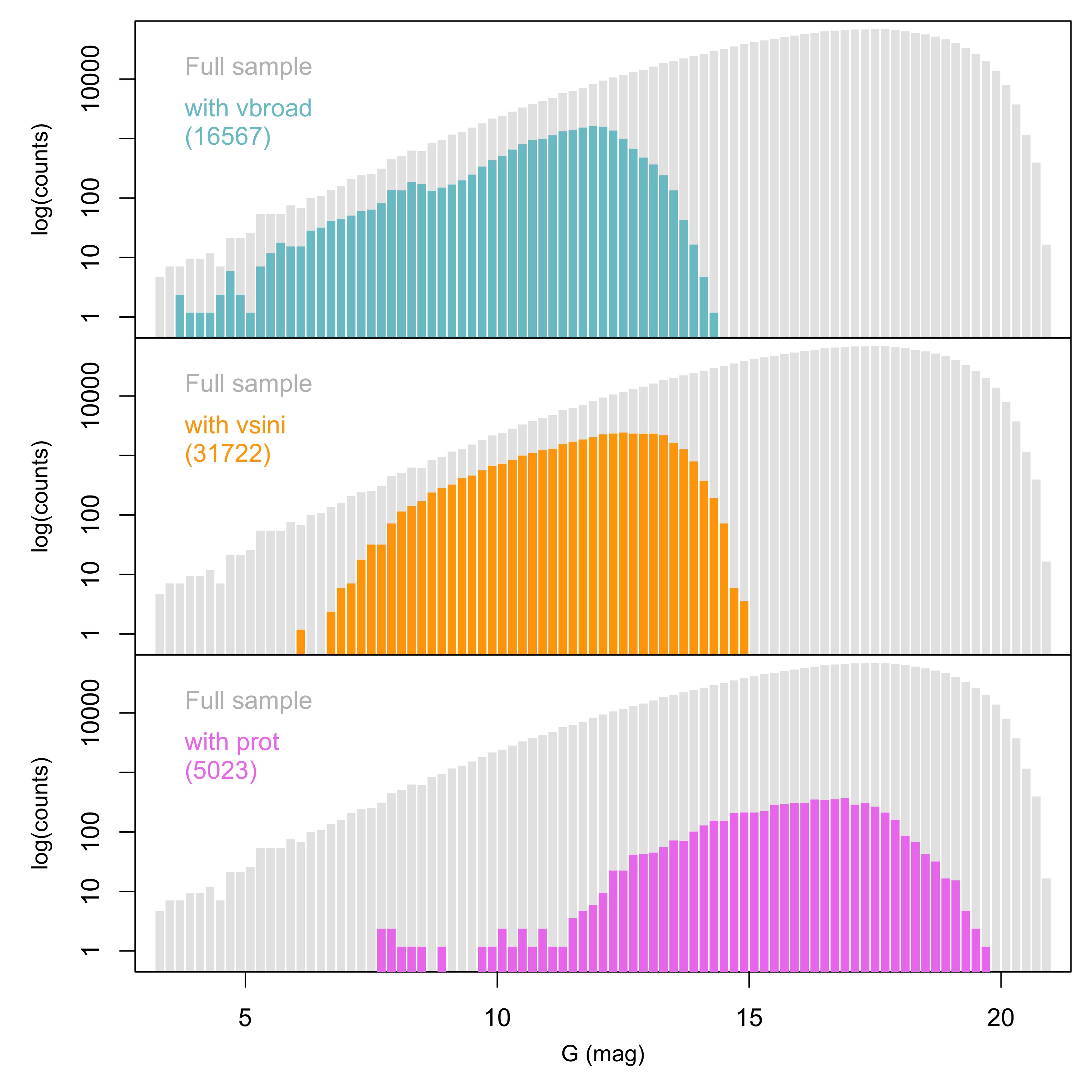}
	\vspace{-0.5cm}
    \caption{Apparent G magnitude distribution of stars with rotation information. In all panels, the full star masterlist is plotted in grey in the background. Stars with a valid {\tt vbroad} parameter are plotted in cyan in the top panel, stars with a valid {\tt vsini} in orange in the middle panel, and stars with a valid {\tt prot} parameter in magenta in the bottom panel (see Secs.~\ref{sec:vbroad}, \ref{sec:vsini}, and \ref{sec:prot} for more details). Sample sizes are listed in parenthesis in each panel.}
   \label{fig:hist}
\end{figure}
% --------------------------------------------

%%%%%%%%%%%%%%%%%%%%%%%%%%%%%%%%%%%%%%%%%%%%%%%%%%

\subsection{Rotational broadening from Gaia RVS spectra}
\label{sec:vbroad}

In {\em Gaia} DR3 the rotational broadening was measured for the first time \citep{fremat22} and is stored in the parameter {\tt vbroad} in the {\tt gaia\_source} Table. While there are still some limitations in these measurements, the large sample size and full sky coverage render the catalogue unique and improvements are expected in the next {\em Gaia} data releases. Briefly, {\tt vbroad} was estimated by cross-correlation of the RVS spectra with broadened synthetic spectra, without correcting for macro-turbulence broadening\footnote{Macroturbulence broadening is expected to be about a few km\,s$^{-1}$, depending on the type of star \citep{doyle14,johnsson20}.}, and filtering rules were applied to the results: {\em (i)} sources with at least 6 transits were considered; {\em (ii)} measurements outside the range 5\,$<$\,v$_{\rm{broad}}\,<$500\,km\,s$^{-1}$ were filtered out; {\em (iii)} only spectra with valid RV measurements were considered, i.e. with 3500\,$<$\,\teff\,$<$14\,500~K; and {\em (iv)} only stars with \grvs\,$<$\,12~mag were analyzed, corresponding to G\,$\simeq$\,14.5~mag (Figure~\ref{fig:hist}).

% --------------------------------------------
\begin{figure}
    \centering
	\includegraphics[width=\columnwidth]{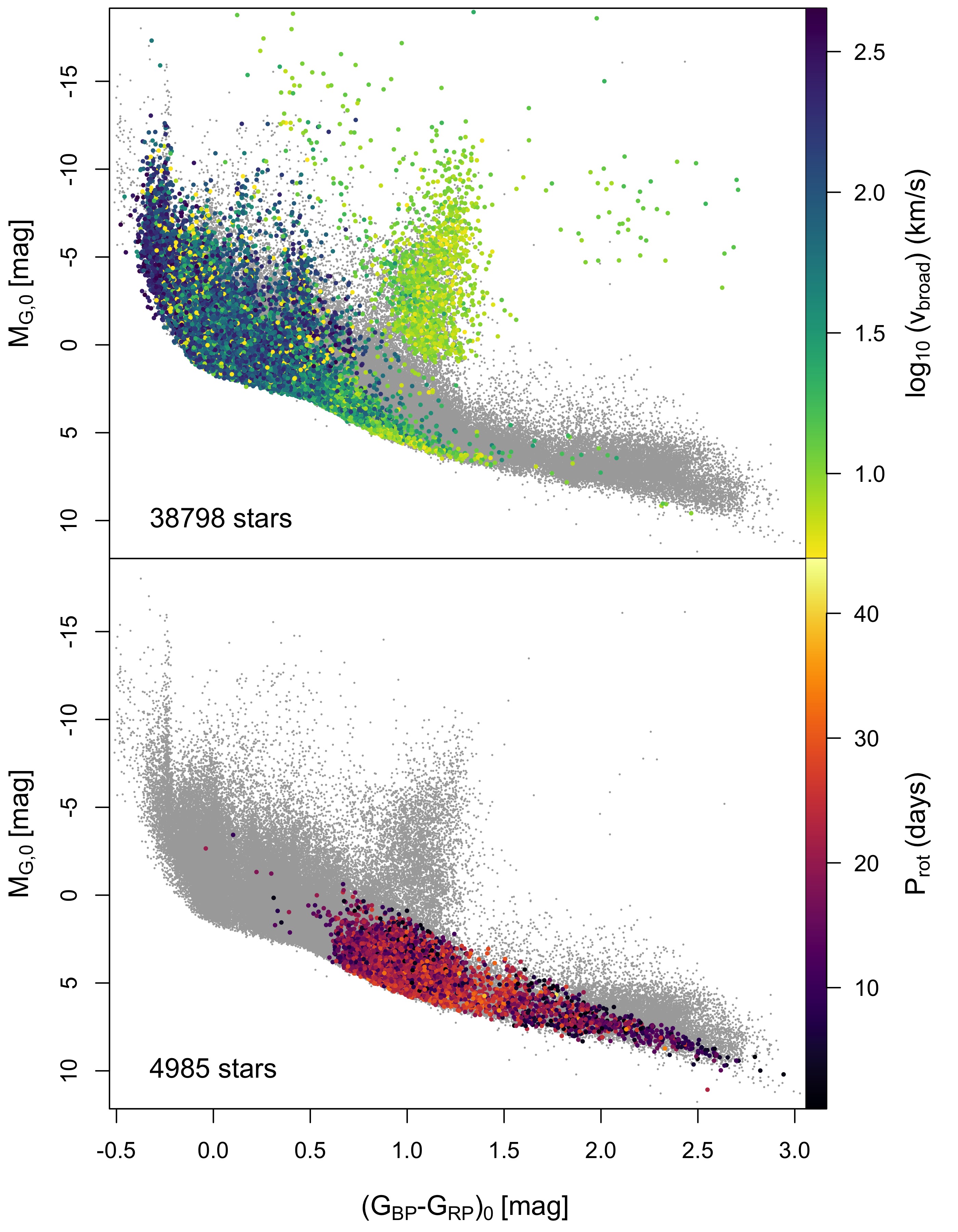}
	\vspace{-0.3cm}
    \caption{Stars with rotation information in {\em Gaia} DR3, on the absolute and dereddened CMD. Stars with a {\tt vbroad} or {\tt vsini} determination are shown on the top panel, while stars with a {\tt prot} determination in the bottom panel. A selection of well-measured stars from the full sample is plotted in grey in the background, for reference.}
   \label{fig:sample}
\end{figure}
% --------------------------------------------

A very careful assessment of the reliability of the DR3 {\tt vbroad} measurements was performed with dedicated simulations by \citet{fremat22} as a function of \teff, spectral mismatches, \grvs, and the like, as well as extensive comparisons with literature samples, including large spectroscopic surveys. As a result, it appears that {\tt vbroad} tends to be overestimated when it is lower than 8--20\,km\,s$^{-1}$, depending on \teff\ \citep[see Table~2 by][]{fremat22}. At the same time, there is a bias with apparent \grvs\ magnitude \citep[see Fig.~14 by][]{fremat22}, in the sense that stars with T$_{\rm{eff}} \gtrsim 7500$\,K have a more and more underestimated {\tt vbroad} as they become fainter, where the exact magnitude at which this becomes important changes as a function of T$_{\rm{eff}}$ \citep[see also Fig.~\ref{fig:surveys} and][]{babusiaux22}. 

%%%%%%%%%%%%%%%%%%%%%%%%%%%%%%%%%%%%%%%%%%%%%%%%%%

\subsection{Rotational broadening for stars hotter than 7500\,K}
\label{sec:vsini}

The {\em Gaia} extended stellar parametrisers (ESP) are data analysis modules used in {\em Gaia} data processing to derive stellar parameters for specific stellar types \citep[see][and the {\em Gaia} online documentation, Sect.~11.3.8\footnote{\url{https://gea.esac.esa.int/archive/documentation/GDR3/Data\_analysis/chap\_cu8par/sec\_cu8par_apsis/ssec\_cu8par\_apsis_esphs.html}}]{fouesneau22}. In particular, the ESP dedicated to hot O, B, and A stars (ESP-HS) uses simultaneously the BP/RP and RVS spectra, when available, to derive stellar parameters by comparison with synthetic spectra, specifically suited for these stars. Among them, ESP-HS also derives the projected rotation ({\tt vsini\_esphs}, hereafter {\tt vsini}), in case the star has RVS spectra. The results are much more scattered than for {\tt vbroad}, as can be appreciated in Fig.~\ref{fig:surveys} (bottom panels), and as expected in the case of hot stars. However, when we compare the typical uncertainties reported in the {\em Gaia} DR3 catalogue for {\tt vbroad} and {\tt vsini}, we find the latter to be about a
factor of two smaller. We thus multiplied the {\tt vsini} uncertainties by a factor of four before using them in the following analysis. The G magnitude distribution of stars with a {\tt vsini} estimate is shown in Fig.~\ref{fig:hist} (middle panel). We observed that {\tt vsini} shows the same trend of {\tt vbroad} in being underestimated for fainter and hotter stars \citep[see also][]{fremat24}. Thus, in this work, we mostly focus on {\tt vbroad} rather than {\tt vsini}, unless specifically indicated.

% --------------------------------------------
\begin{figure}
    \centering
	\includegraphics[width=\columnwidth]{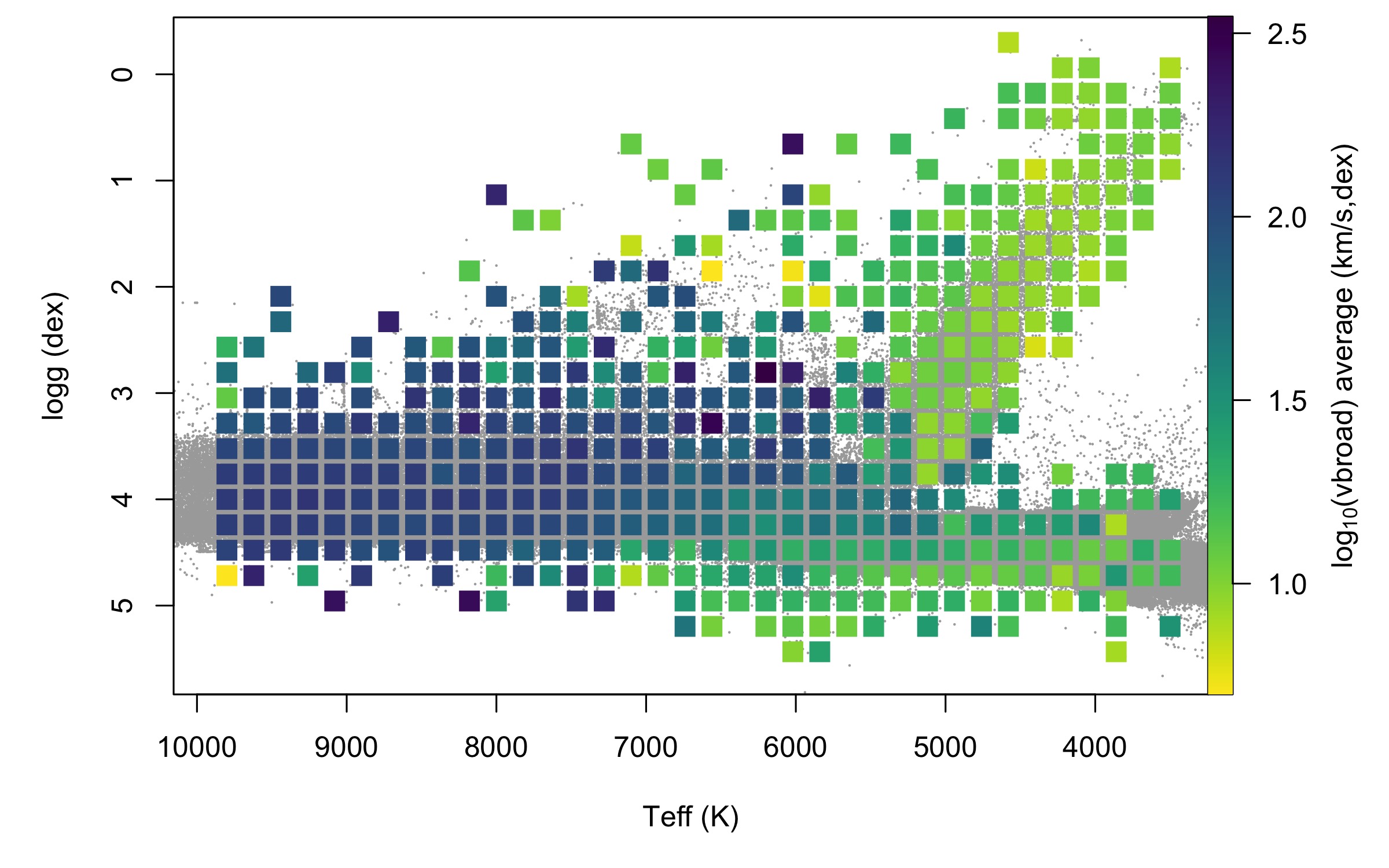}
	\vspace{-0.3cm}
    \caption{The average of {\tt vbroad} or, when missing, {\tt vsini}, in a 2D set of bins in the Kiel diagram. The colorscale is logarithmic to make any subtle variations more visible. In the background, the entire sample is plotted in grey. We left out stars hotter than 10\,000\,K, we limited the colorscale between 5 and 400\,km\,s$^{-1}$, and we only plotted bins with at least three stars, for clarity.}
   \label{fig:frac1}
\end{figure}
% --------------------------------------------
% --------------------------------------------
\begin{figure}
    \centering
	\includegraphics[width=\columnwidth]{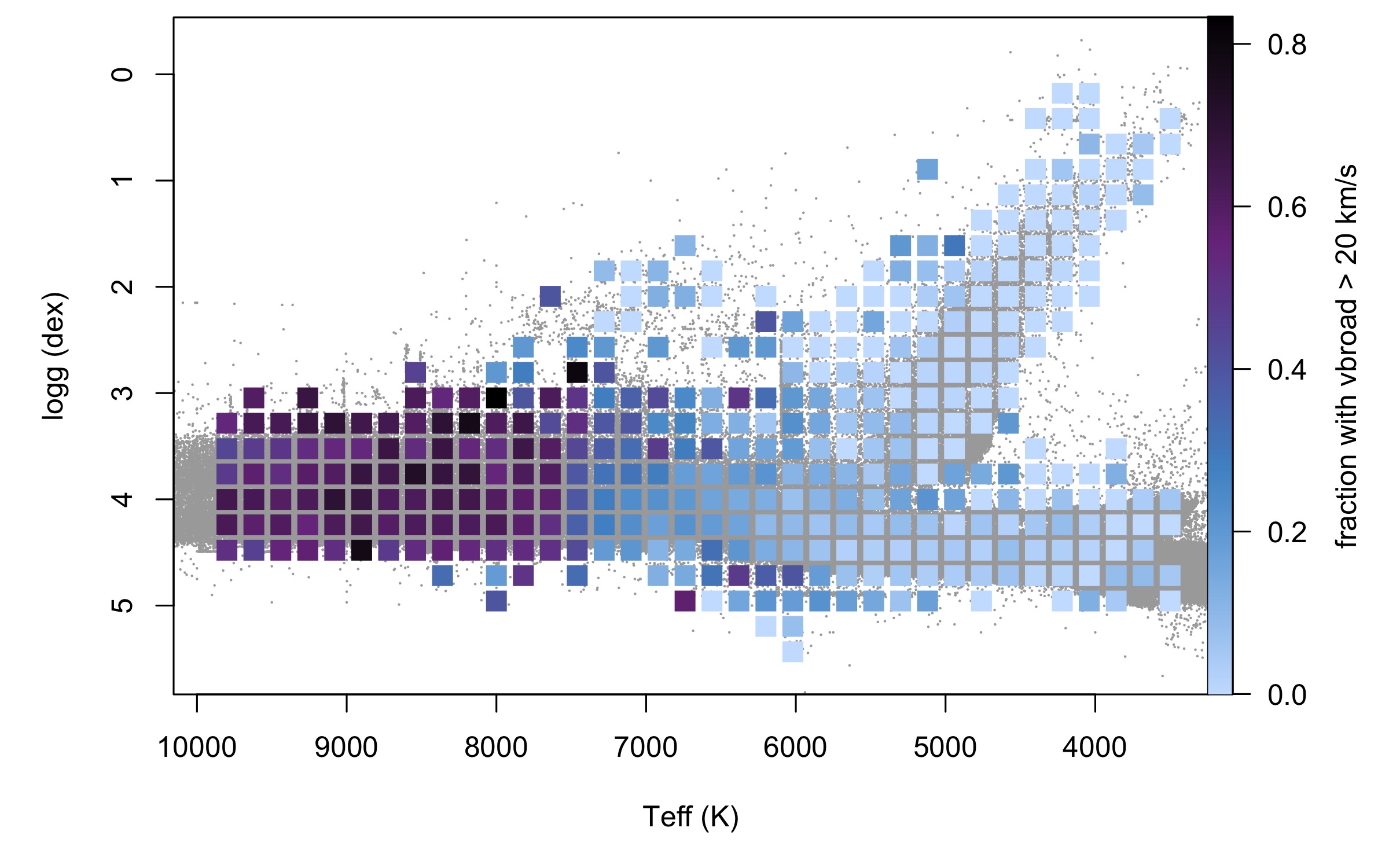}
	\vspace{-0.3cm}
    \caption{Similar to Fig.~\ref{fig:frac1}, but here we plot the fraction of stars with {\tt vbroad}\,$>$\,20\,km\,s$^{-1}$. In the background, the entire sample is plotted in grey. We left out stars hotter than 10\,000\,K and fainter than G\,=\,14\,mag (which could not possibly have a {\tt vbroad} estimate in {\em Gaia} DR3). We only plotted bins with at least five stars, for clarity.}
   \label{fig:frac2}
\end{figure}
% --------------------------------------------

%%%%%%%%%%%%%%%%%%%%%%%%%%%%%%%%%%%%%%%%%%%%%%%%%%

\subsection{Rotational periods from Gaia epoch photometry}
\label{sec:prot}

The {\em Gaia} multi-epoch photometric data allow for the detailed study of several types of variable stars \citep{eyer22}. The determination of rotational periods ({\tt best\_rotation\_period}, hereafter {\tt prot}) in {\em Gaia} DR2 and DR3 is based on the flux variability induced by asymmetrically distributed spots on the stellar surface and is only computed for stars of relatively late spectral types in the main sequence \citep[\teff$\lesssim$6\,500~K,][]{lanzafame18,distefano22}. This choice has the additional advantage of minimizing contamination by stellar pulsators. The information is stored in the {\tt vari\_rotation\_modulation} Table in the {\em Gaia} archive. An estimate of the chromospheric activity was also obtained as the flux variation amplitude, A$_{\rm{max}}$ ({\tt max\_activity\_index\_g}). Interestingly, the DR2 sample is not fully included in the DR3 one, thus we downloaded both catalogues. We used our cross-match between DR2 and DR3, described above, to assign a DR3 source ID to sources in the DR2 rotation modulation catalogue. We then merged the two catalogues. When a source was present in both DR2 and DR3, we selected the DR3 information. 
The G magnitude distribution of stars with a {\tt prot} estimate is shown in Fig.~\ref{fig:hist} (bottom panel). As can be seen, these stars are generally much fainter than the ones having {\tt vbroad} or {\tt vsini}, providing an entirely different window on stellar rotation.

% --------------------------------------------
\begin{figure*}
    \centering
	\includegraphics[width=\textwidth]{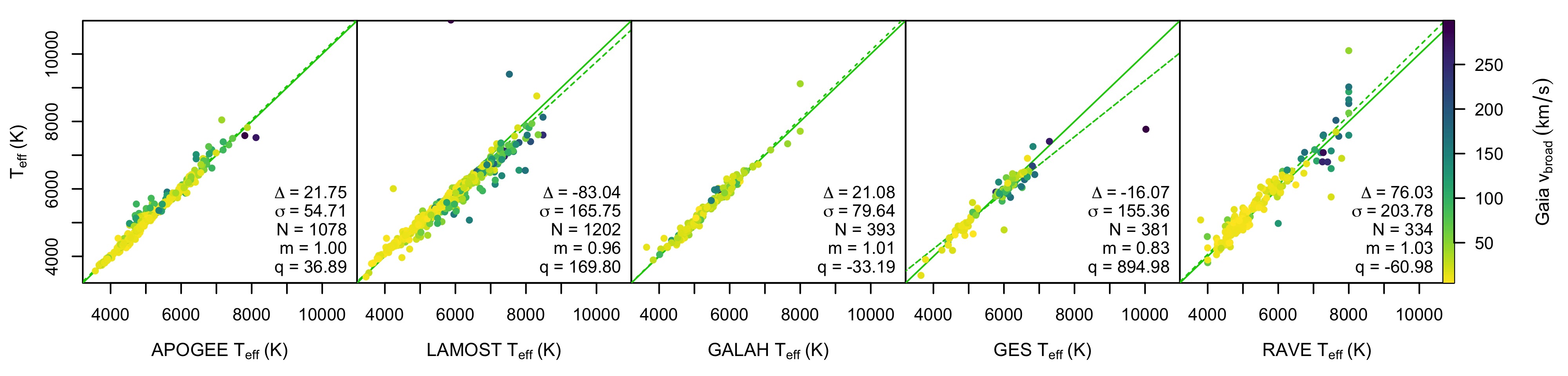}
	\includegraphics[width=\textwidth]{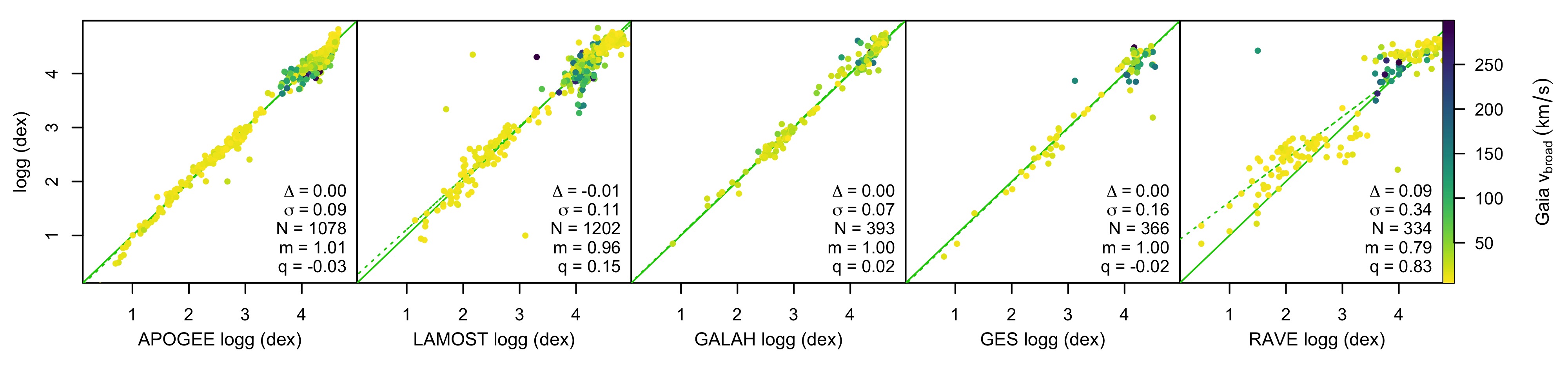}
	\vspace{-0.5cm}
    \caption{Comparison of our selected \teff\ (top row) and log$g$ (bottom row) parameters with those of the main spectroscopic surveys (see Section~\ref{sec:vbroad} for more details). Each panel reports a different survey as well as the one to one relation (solid green lines) and a linear fit (dashed green lines). The points are colored based on their {\em Gaia} {\tt vbroad}, where the scale saturates at 300\,km\,s$^{-1}$ to increase plot clarity. The statistics of the comparison and the parameters of linear fits are reported in each panel, labelled as in Figure~\ref{fig:surveys}.}
   \label{fig:pars}
\end{figure*}
% --------------------------------------------

A careful validation of the {\tt prot} estimates was performed by \citet[][for DR2]{lanzafame18} and \citet[][for DR3]{distefano22}, which results in the cleaning of stars with non-optimal time sampling or coverage, in the cleaning of spurious periods, and in a generally favorable comparison with literature samples such as the Zwicky Transient Facility survey \citep[ZTF,][]{chen20}, the ASAS-SN survey \citep{jayasinghe21}, and several other high-quality literature studies. 
We note that the particular time sampling of {\em Gaia} does not allow for a reliable determination of periods larger than a few tens of days. As in the case of {\tt vbroad} and {\tt vsini}, the major strength of the {\tt prot} catalogue lies in its unprecedented size and full-sky coverage, which for example allowed for the definition of a new class of stellar rotators \citep[][see also Section~\ref{sec:lowMS}]{lanzafame18,lanzafame19}. 

%%%%%%%%%%%%%%%%%%%%%%%%%%%%%%%%%%%%%%%%%%%%%%%%%%

\section{Sample characterization details}
\label{app:params}

%%%%%%%%%%%%%%%%%%%%%%%%%%%%%%%%%%%%%%%%%%%%%%%%%%

\subsection{Stellar radial velocities}
\label{sec:rvst}

We compared the {\em Gaia} DR3 RVs with the ones in the Survey of Surveys \citep[SoS,][]{tsantaki22}, which are obtained by homogenizing, merging, and recalibrating with high-quality samples all the RV measurements by the major spectroscopic surveys. In particular: {\em Gaia} DR2 \citep{gdr2}; APOGEE DR16 \citep{ahumada20}\footnote{\url{https://www.sdss.org/dr16/}}; RAVE DR6 \citep{steinmetz20}\footnote{\url{https://www.rave-survey.org/}}; GALAH DR2 \citep{buder18}\footnote{\url{https://www.galah-survey.org/}}; LAMOST DR5 \citep{zhao12}\footnote{\url{http://www.lamost.org/public/}}; and Gaia-ESO DR3 \citep{gilmore12}\footnote{\url{https://www.gaia-eso.eu/}}. No trends are apparent in the DR3 RVs with any of the parameters considered by \citet{tsantaki22}, while the G magnitude trend previously reported for the DR2 RVs by \citet{katz19} and \citet{tsantaki22} is more than halved in DR3, reaching a maximum overestimate of about 0.4\,km\,s$^{-1}$ at G$_{\rm{RVS}}$\,$\simeq$\,14\,mag. We corrected the trend using Equation~5 by \citet{katz22}. A different trend was observed for stars hotter than 8500\,K, which we corrected using Equation~1 by \citet{blomme22}. 

After these corrections, the median difference of the {\em Gaia} RVs from the SoS ones is --0.06\,$\pm$\,2.03\,km\,$^{-1}$ for the full sample. As expected, the agreement worsens when considering only stars with {\tt vbroad}\,$>$\,50\,km\,s$^{-1}$, where it becomes --1.56\,$\pm$\,6.67\,km\,s$^{-1}$. In any case, the comparison is compatible with the typical precision and accuracy quoted in the respective papers, therefore we adopt the SoS value for $\simeq$6700 faint stars without a {\em Gaia} RV estimate in DR3. Finally, we added RVs from the latest versions of the surveys, after correcting for the median zero-point with respect to {\em Gaia} DR3. This way, we could assign a radial velocity estimate to $\simeq$14\% of our sample stars. Because our collection of individual RV estimates is larger than the one by \citet{hunt24}, we used it to determine new cluster mean velocities (Sect.~\ref{sec:ocs}). For clusters with at least 10 stars, we also estimated an RV membership, listed in the {\tt RV\_mem} parameter in Table~\ref{tab:stars}.

%%%%%%%%%%%%%%%%%%%%%%%%%%%%%%%%%%%%%%%%%%%%%%%%%%

\subsection{Stellar parameters}
\label{sec:stpars}

We used several literature source catalogues, prioritised by their precision, which were estimated with the three-cornered hat method, using the stars in common between pairs of catalogues. We noted that the median differences (i.e. zeropoints) among the quoted catalogues are typically below 100\,K in \teff, 0.2\,dex in log\,$g$, and 0.3 dex in [Fe/H] and they get considerably better (about halved) when considering stars with 4000\,$<$\,\teff\,$<$\,7500\,K. We also noted that the spreads in the pairwise comparisons among catalogues are in some cases larger than the sum in quadrature of the quoted uncertainties. This is not a problem for the purpose of the present work, but we stress that the reported uncertainties may be underestimated and should be used with caution. The catalogues we used, in priority order, to assign \teff, log\,$g$, and [Fe/H] estimates, are: 
\begin{itemize}
\item{SoS spectroscopic parameters \citep{turchi25};}
\item{spectroscopic surveys parameters from the latest available releases (see Sect.~\ref{sec:rot});}
\item{{\em Gaia} DR3 spectroscopic parameters  from {\tt gspspec}, after applying the recommended corrections to metallicity and log\,$g$ \citep{recio22}, and excluding stars with metallicity lower than --1\,dex;}
\item{SoS machine-learning parameters \citep{turchi25};}
\item{machine learning [Fe/H] from \citet{andrae23}, where we only considered stars with \,\teff\,$<$\,4000\,K and we did not use their \teff\ and log\,$g$ parameters;}
\item{{\rm Gaia} DR3 photometric parameters from {\tt gspphot}, excluding stars with metallicity lower than --1\,dex;}
\item{{\rm Gaia} DR3 ESPHS parameters for stars hotter than 7\,500\,K.}
\end{itemize}

\noindent This way, we could assign at least a rough estimate of \teff\ and of log\,$g$ to $\simeq$49\% of our sample stars, and an estimate of [Fe/H] to $\simeq$45\% of the sample stars. We note that the typical uncertainties of this collection of parameters are of the order of about 200--500\,K  in \teff, 0.2--0.5\, dex in log\,$g$, and 0.2--0.5\,dex in [Fe/H], with large variations within the sample across the parameter space. In Fig.~\ref{fig:pars} we compare our collected parameters with the latest versions of the main spectroscopic surveys (see Sect.~\ref{sec:rot}). As can be seen, there is a good overall agreement, except for stars hotter than about 8000\,K or with large {\tt vbroad}.

% --------------------------------------------
\begin{figure}
    \centering
	\includegraphics[width=\columnwidth]{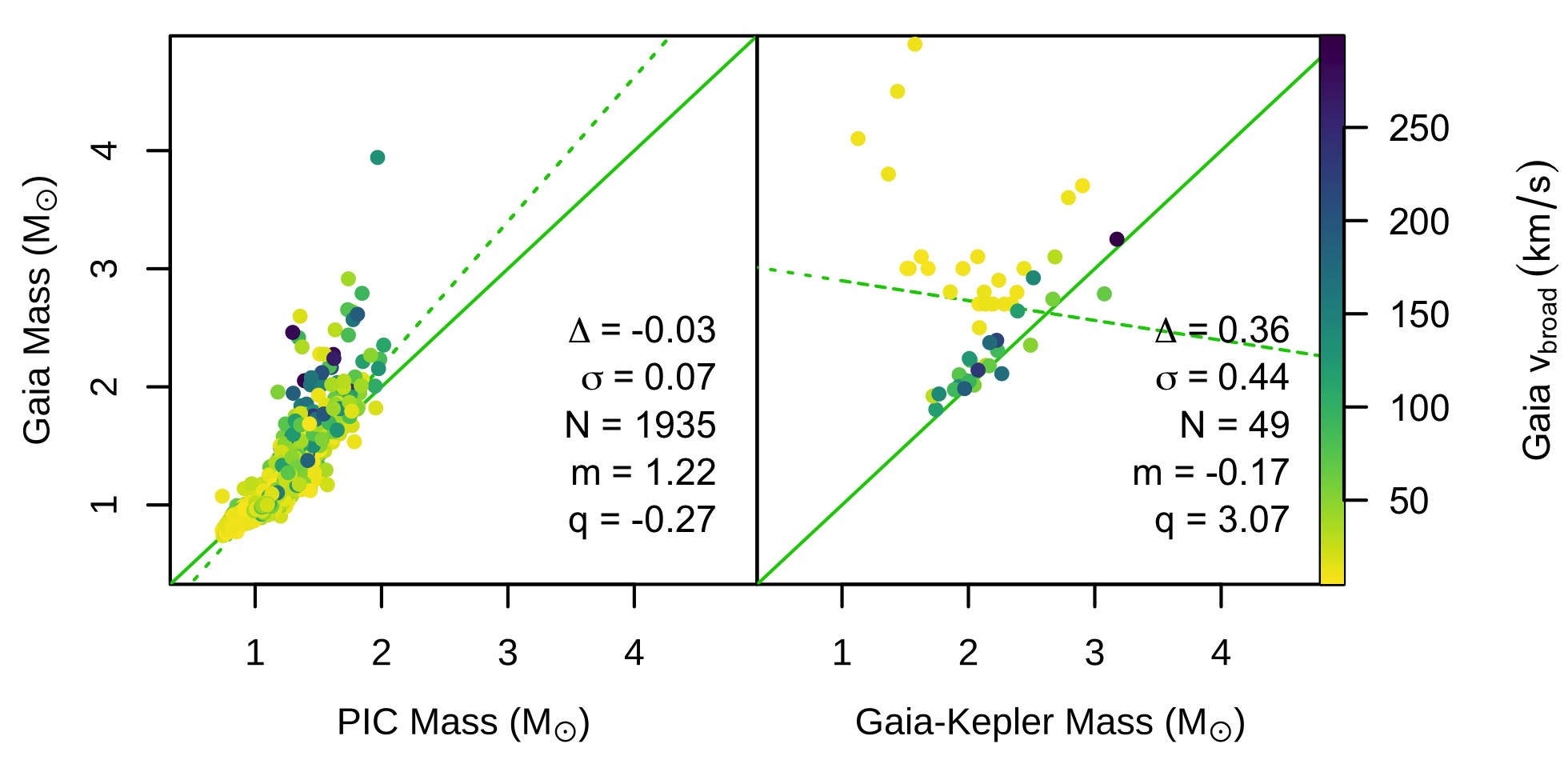}
	\includegraphics[width=\columnwidth]{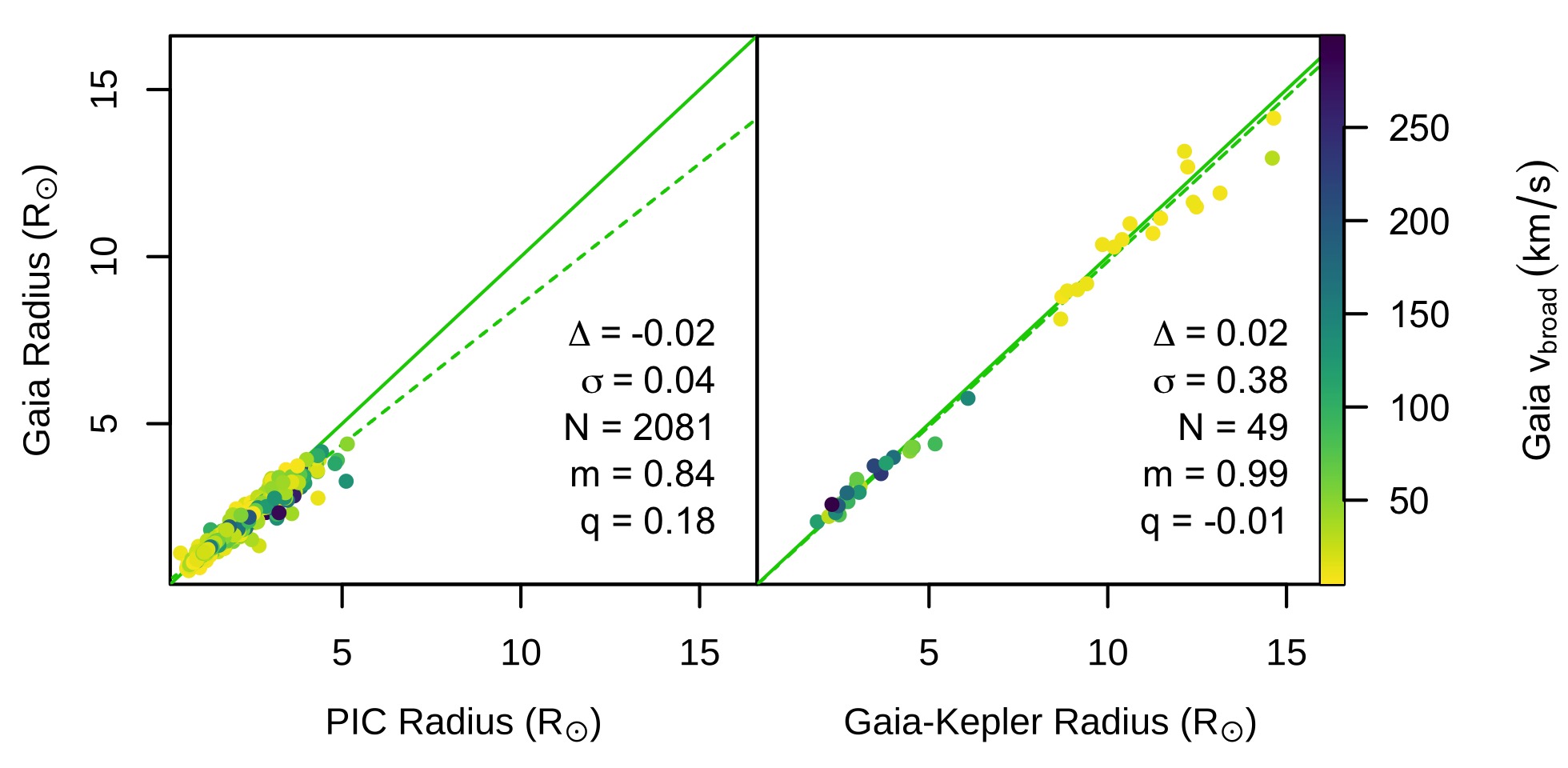}
	\vspace{-0.5cm}
    \caption{Similar to Figure~\ref{fig:pars}, but for the case of mass (top panels) and radius (bottom panels) estimates. The left panels show the comparison with the PLATO Input Catalogue \citep[PIC,][]{montalto21}, and the right ones with the Gaia-Kepler catalogue \citep{berger20}. The symbols, annotations, and color-scale are the same as in Figure~\ref{fig:pars}.}
   \label{fig:kep}
\end{figure}
% --------------------------------------------

For stellar masses and radii, we used the {\em Gaia} DR3 FLAME results (Final Luminosity Age and Mass Estimator), which were extensively tested \citep{babusiaux22}, showing some problems only in the high mass range. This is confirmed by our comparison with the PLATO input catalogue \citep{montalto21} and the Gaia-Kepler catalogue \citep{berger20}, as illustrated in Fig.~\ref{fig:kep}. However, for masses lower than about 2.5\,M$_{\rm{\odot}}$ the different datasets agree with each other more than satisfactorily, as well as for radii. We found only mild indications that the {\em Gaia} FLAME parameters are worse for stars with larger {\tt vbroad} (Fig.~\ref{fig:kep}, top-right panel).

%%%%%%%%%%%%%%%%%%%%%%%%%%%%%%%%%%%%%%%%%%%%%%%%%%

\subsection{Cluster properties}
\label{sec:ocs}

We compared the distance estimates by \citet{hunt24} against the median of the \citet{bailer21} distances and against the values in our literature collection (Sect.~\ref{sec:starmaster}). We noticed that after about 5\,kpc the agreement among these catalogues breaks down and the spread rises to 10-15\%, respectively. We attribute this to a parallax bias between {\em Gaia} DR2 and DR3 of about 0.015~mas (estimated from the two member lists). Indeed the \citet{bailer21} median cluster distances agree better with the \citet{hunt24} ones than with our literature collection, which is dominated by DR2 estimates. We thus decided to rely on the \citet{hunt24} estimates and we used the \citet{bailer21} ones for the 51 clusters not included in their catalogue. Cluster masses are taken from \citet{hunt24}.

For mean RVs, we profited from the values collected in our member stars list (Sect.~\ref{sec:stpars} and Table~\ref{tab:stars}), which come mostly from {\em Gaia} DR3 \citep{katz22,blomme22} and the SoS \citep{tsantaki22}, but also from spectroscopic surveys. This way, we can count on more stars with RV estimated than the ones considered by \citet{hunt24}. We computed cluster median RVs and their spread using the median and the MAD of the members stars in each cluster, after removing all the confirmed binary and variable stars. Our results agree extremely well with the ones by \citet{hunt24}, with a median difference of -0.001\,$\pm$\,3.249\,km\,s$^{-1}$, but we could add RV to almost 500 more clusters and more stars per cluster are generally used to obtain each estimate. We included the RV estimates by \citet{dias21} for four clusters lacking it. In summary, we obtained a systemic RV estimate for 4980 clusters ($\simeq$\,87\%), but we note that 2064 clusters have $<$3 stars with a valid RV estimate.

Cluster metallicities are not present in the \citet{hunt21,hunt23,hunt24} catalogue. We thus collected literature results by \citet{dias21}, \citet{netopil16,netopil22}, \citet{fu22}, \citet{zhang24} for 1571 clusters. We also computed new cluster metallicities from our collection of [Fe/H] estimates, by selecting stars with --1\,$<$\,[Fe/H]\,$<$\,0.5\,dex, \teff\,$<$8000\,K, and with metallicity errors below 1\,dex. When comparing the literature estimates with each other we obtained values within less than 0.05--0.10\,dex from each other. When comparing our newly computed mean cluster metallicities with the literature collection, however, we found offsets of 0.2\,dex or more and a very poor correlation. Only once the {\em Gaia} DR3 individual star estimates were removed from the collection, we could obtain estimates comparable to the ones in the literature, with a difference of $\Delta$[Fe/H]\,=\,0.03\,$\pm$\,0.13\,dex for 980 clusters in common. Merging the two sets, the literature one and our newly computed metallicites, we could assign an average [Fe/H] estimate to 2336 clusters in our sample (about 41\%).

Finally, for cluster ages, we noticed that for relatively old clusters ($\gtrsim$500\,Myr) with many BSS, the ages by \citet{hunt23} are severely underestimated, because the isochrone is wrongly adjusted to include the BSS sequence. This is particular evident in their Fig.~6 for the case of Ruprecht\,147. For this reason, we preferred the ages by \citet{cavallo24} when available. Otherwise, we used \citet{cantat20_250k} because of the good agreeement with \citet{cavallo24}, and in case those were missing, we used \citet{hunt23}.

%%%%%%%%%%%%%%%%%%%%%%%%%%%%%%%%%%%%%%%%%%%%%%%%%%
\section{Variable and binary stars}
\label{app:varbin}

% --------------------------------------------
\begin{figure}
    \centering
\includegraphics[width=\columnwidth]{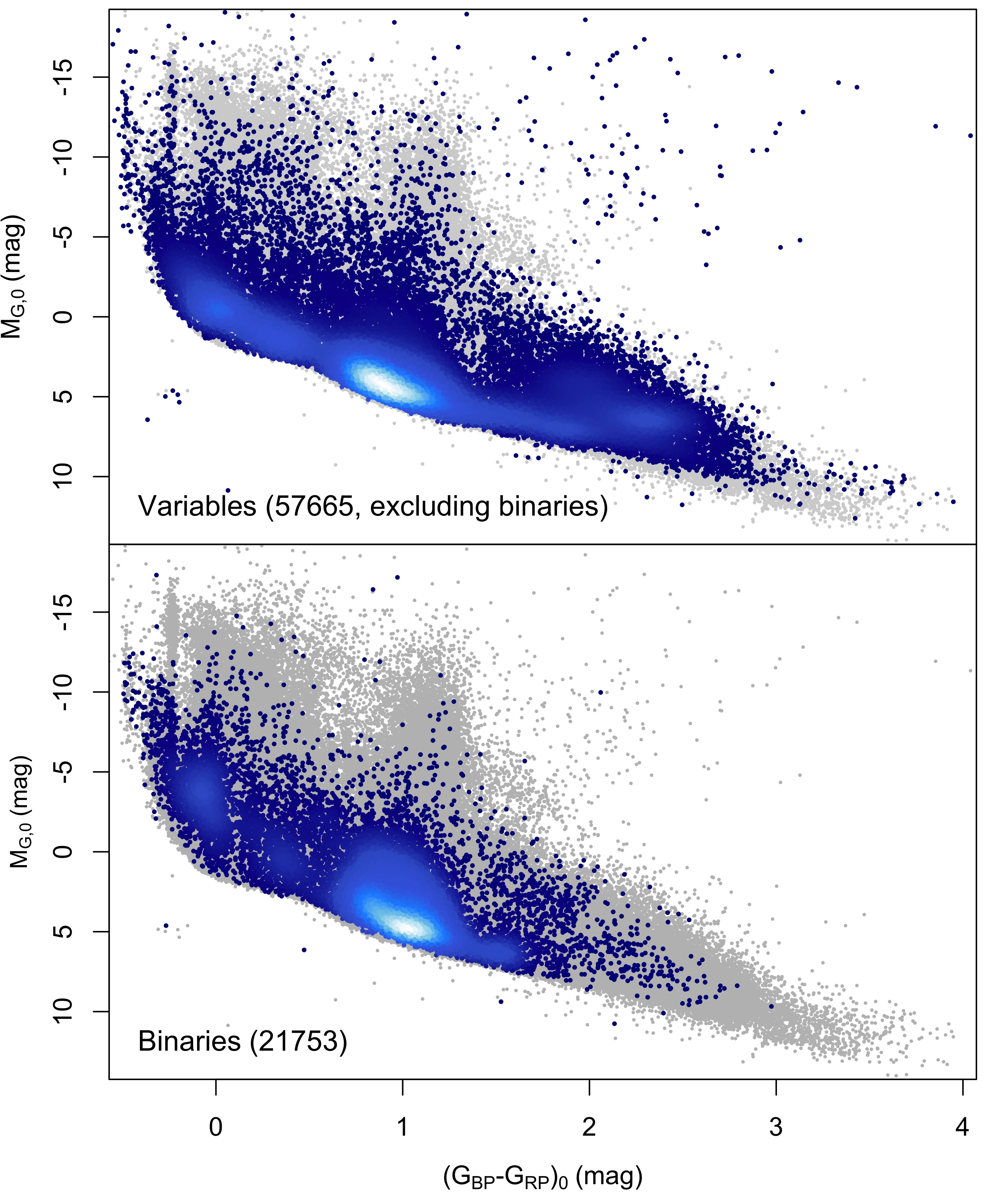}
%\vspace{-0.3cm}
    \caption{Density of variable stars (top panel, excluding binaries) and binary stars (bottom panel) in the absolute and dereddened CMD. Both confirmed and candidate variables and binaries are included. The full sample (Table~\ref{tab:stars}) is plotted in the background as small grey points in both panels.}
   \label{fig:varbin}
\end{figure}
% --------------------------------------------

% ------------------------------------------------
\begin{table}
	\centering
	\caption{Literature information on variable stars. See Sects.~\ref{sec:var}, \ref{sec:bin}, and \ref{app:varbin} for more details and references.}
	\label{tab:varinfo}
	\begin{tabular}{lll} 
		\hline
		Quantity    & Units & Description \\
		\hline
        source\_id  &       & {\em Gaia} DR3 source ID \\
        GAIA\_PER   & (d)   & {\em Gaia} DR3 period \\
        GAIA\_AMP   & (mag) & {\em Gaia} DR3 amplitude \\
        GAIA\_NUM   &       & {\em Gaia} DR3 epochs \\
        GAIA\_TYP   &       & {\em Gaia} DR3 type \\
        GAIA\_CLASS &       & {\em Gaia} DR3 class \\
        GAIA\_SCORE &       & {\em Gaia} DR3 score \\
        GAV\_PER    & (d)   & \citet{gavras23} period \\
        GAV\_TYP    &       & \citet{gavras23} superclass \\
        GAV\_SRC    &       & \citet{gavras23} source \\
        ZTF\_ID     &       & ZTF identifier \\
        ZTF\_PER    & (d)   & ZTF period \\
        ZTF\_AMP    & (mag) & ZTF amplitude \\
        ZTF\_NUM    &       & ZTF epochs \\
        ZTF\_TYP    &       & ZTF classification \\
        ASA\_ID     &       & ASAS-SN identifier \\
        ASA\_PER    & (d)   & ASAS-SN period \\
        ASA\_AMP    & (mag) & ASAS-SN amplitude \\
        ASA\_PROB   &       & ASAS-SN class probability \\
        ASA\_TYP    &       & ASAS-SN classification \\
        TES\_ID     &       & TESS identifier \\
        TES\_PER    & (d)   & TESS period \\
        TES\_AMP    & (mag) & TESS amplitude \\
        TES\_PROB   &       & TESS class probability \\
        TES\_TYP    &       & TESS classification \\
        OGL\_ID     &       & OGLE identifier \\
        OGL\_PER    & (d)   & OGLE period \\
        OGL\_AMP    & (mag) & OGLE amplitude \\
        OGL\_TYP    &       & OGLE classification \\       
		\hline
	\end{tabular}
\end{table}
% ------------------------------------------------

% ------------------------------------------------
\begin{table}
	\centering
	\caption{Literature information on binaries or suspected binary stars. See Sects.~\ref{sec:var}, \ref{sec:bin}, and \ref{app:varbin} for more details and references.}
	\label{tab:bininfo}
	\begin{tabular}{lll} 
		\hline
		Quantity    & Units & Description \\
		\hline
        source\_id  &       & {\em Gaia} DR3 source ID \\
        NSS\_PER    & (d)   & {\em Gaia} NSS orbital period \\
        NSS\_ECC    &       & {\em Gaia} NSS orbital eccentricity \\
        NSS\_TYP    &       & {\em Gaia} NSS solution type \\
        SB9\_ID     &       & SB9 identifier \\
        SB9\_PER    & (d)   & SB9 orbital period \\
        SB9\_ECC    &       & SB9 orbital eccentricity \\
        SB9\_TYP    &       & SB9 binary type \\
        SB9\_BIB    &       & SB9 reference \\
        KIR\_ID     &       & \citet{kirk16} identifier \\
        KIR\_TYP    &       & \citet{kirk16} binary type \\
        DEL\_ID     &       & \citet{deleuil18} identifier \\
        DEL\_PER    & (d)   & \citet{deleuil18} orbital period \\
        DEL\_TYP    &       & \citet{deleuil18} binary type \\
        ELB\_ID     &       & \citet{elbadry18} identifier \\
        ELB\_TYP    &       & \citet{elbadry18} binary type \\
        BIR\_ID     &       & \citet{birko19} identifier \\
        BIR\_TYP    &       & \citet{birko19} binary type \\
        QIA\_ID     &       & \citet{qian19} identifier \\
        QIA\_TYP    &       & \citet{qian19} binary type \\
        MAZ\_ID     &       & \citet{mazzola20} identifier \\
        MAZ\_TYP    &       & \citet{mazzola20} binary type \\
        PRI\_ID     &       & \citet{price20} identifier \\
        PRI\_NUM    &       & \citet{price20} epochs \\
        PRI\_PER    & (d)   & \citet{price20} period \\
        PRI\_ECC    &       & \citet{price20} eccentricity \\
        PRI\_TYP    &       & \citet{price20} binary type \\
        TIA\_ID     &       & \citet{tian20} identifier \\
        TIA\_PROB   &       & \citet{tian20} epochs \\
        TIA\_TYP    &       & \citet{tian20} binary type \\
        TRA\_ID     &       & \citet{traven20} identifier \\
        TRA\_TYP    &       & \citet{traven20} binary type \\
        KOU\_ID     &       & \citet{kounkel21} identifier \\
        KOU\_NUM    &       & \citet{kounkel21} epochs \\
        KOU\_PER    & (d)   & \citet{kounkel21} period \\
        KOU\_ECC    &       & \citet{kounkel21} eccentricity \\
        KOU\_TYP    &       & \citet{kounkel21} binary type \\
        MER\_ID     &       & \citet{merle17,merle20} identifier \\
        MER\_TYP    &       & \citet{merle17,merle20} binary type \\
        KOV\_ID     &       & \citet{kovalev24} identifier \\
        KOV\_PER    & (d)   & \citet{kovalev24} period \\
        KOV\_ECC    &       & \citet{kovalev24} eccentricity \\
        KOV\_TYP    &       & \citet{kovalev24} binary type \\
        GRO\_ID     &       & \citet{grondin24} identifier \\
        GRO\_TYP    &       & \citet{grondin24} binary type \\
        JAC\_ID     &       & \citet{jackim24} identifier \\
        JAC\_TYP    &       & \citet{jackim24} binary type \\
		\hline
	\end{tabular}
\end{table}
% ------------------------------------------------ 

%%%%%%%%%%%%%%%%%%%%%%%%%%%%%%%%%%%%%%%%%%%%%%%%%%

\subsection{Variable stars}
\label{sec:var}

Variability information was obtained mainly from the {\em Gaia} DR3 {\tt vari\_summary} and {\tt vari\_classifier\_result} tables. The enormous amount of work done to classify and characterise the {\em Gaia} DR3 variables is summarised by \citet{eyer22} and more details are provided in a series of papers divided by variable type \citep[][]{clementini22,deridder22,distefano22,gomel22,lebzelter22,marton22,ripepi22}. We also profited from the critically compiled catalogue of literature variables by \citet{gavras23}, which contains an exhaustive selection of publications before 2021. We complemented it with the most updated versions of the ZTF survey \citep{chen20}, the ASAS-SN catalogue of variable stars \citep{jayasinghe21}, the latest TESS variables classification by \citet{gao24}. We also checked the updated versions of the main OGLE catalogues\footnote{\url{https://ogledb.astrouw.edu.pl/~ogle/OCVS/}} \citep[][and updates]{pawlak16,soszynski16,pietrukowicz21,soszynski21,iwanek22}. We finally explored the lists of variable objects found in the {\em Gaia} science alerts \citep{hodgkin21,mistry22}, but we only found stars already included in the above catalogues. 

The cross-match of large catalogues with {\em Gaia} DR3 was performed with the software described by \citep{marrese17,marrese19}, while the small ones were cross-matched with {\tt TopCat}\footnote{\url{https://www.star.bristol.ac.uk/mbt/topcat/}} \citep{topcat}. Some literature catalogues were already cross-matched with {\em Gaia} DR3 and in those cases we trusted the provided identifications. We did not attempt any homogenization of the literature information, but we reported some of the basic literature information in Table~\ref{tab:varinfo}. Based on the available literature information, we set a specific flag in the catalogue of cluster members (Table~\ref{tab:stars}), named {\tt flag\_var}, which can take the value {\tt VAR} for confirmed and characterised variable stars, or the value {\tt VAR:} for less secure identifications. To this aim, we employed the available indicators such as figures of merit, classification probability, class identifiers, or conflicting information in the literature. The variable stars sample is presented in Fig.~\ref{fig:varbin} (upper panel).

%%%%%%%%%%%%%%%%%%%%%%%%%%%%%%%%%%%%%%%%%%%%%%%%%%

\subsection{Binary stars}
\label{sec:bin}

{\em Gaia} DR3 presents for the first time the results of the non-single stars pipelines \citep[hereafter NSS,][]{arenou23}, which use astrometric \citep{halbwachs22}, photometric \citep{eyer22}, and spectroscopic time series to detect and to model binary stars \citep[see also][]{holl22,shion22,rimoldini23,mowlavi23}. We use here {\em Gaia} DR3 data for binaries which were observed for at least one entire orbit and modelled as a two body system (table {\tt nss\_two\_body\_orbit}), as well as those for which the orbit was not closed, but a clear variation in the astrometric solution or in the radial velocities was found (tables {\tt nss\_acceleration\_astro} and {\tt nss\_non\_linear\_spectro}).

We complemented the {\em Gaia} binary data with various literature sources, cross-matching them with our sample as described in the previous section. We included binaries from CoRoT \citep{deleuil18} and Kepler \citep{kirk16}, as well as the latest version of the SB9 catalogue \citep[][March 2021 update]{pourbaix04}. We also included binaries from large spectroscopic surveys and other survey data from a variety of catalogues \citep[e.g.][]{merle17,elbadry18,birko19,qian19,merle20,mazzola20,price20,tian20,traven20,kovalev24,grondin24,jackim24}.

Similarly to the case of variable stars, we used the available information in the literature catalogues to evaluate the robustness of the binary identification (i.e. number of epochs and various figures of merit). We labelled binaries with {\tt BIN} and candidate or suspected binaries with {\tt BIN:} in a dedicated field ({\tt flag\_bin}, see Table~\ref{tab:stars}). All the variable stars of binary nature, such as eclipsing or ellipsoidal variables, and many others, were flagged both as variables and as binaries. As in the case of variable stars, we did not attempt to homogenise the literature classification and parametrization of binaries, but we reported relevant literature information in Table~\ref{tab:bininfo}.
Our binary star sample is presented in Fig.~\ref{fig:varbin} (lower panel).

%%%%%%%%%%%%%%%%%%%%%%%%%%%%%%%%%%%%%%%%%%%%%%%%%%

\section{Additional figures and considerations}

%%%%%%%%%%%%%%%%%%%%%%%%%%%%%%%%%%%%%%%%%%%%%%%%%%

% --------------------------------------------
\begin{figure*}
    \centering
    \includegraphics[width=\textwidth]{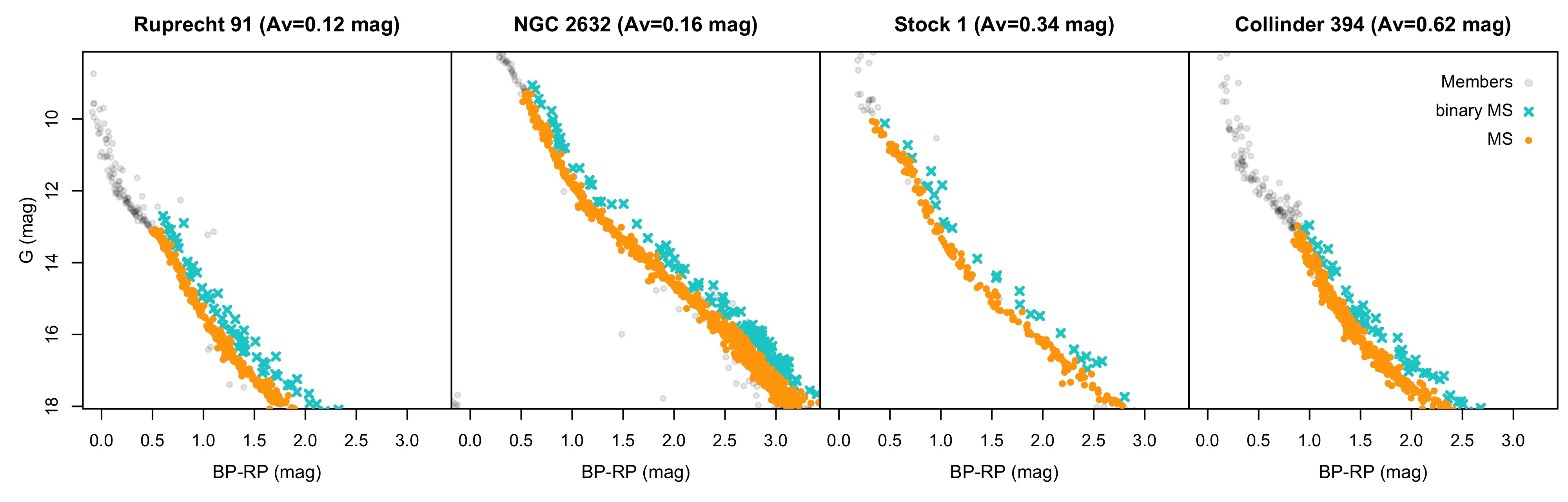}
    \caption{Examples of MS and binMS selections. The clusters are plotted in order of increasing interstellar absorption.}
   \label{fig:binMS}
\end{figure*}
% --------------------------------------------

\subsection{Binary fraction}
\label{sec:binfrac}

We decided not to use our MS and binMS samples to derive an accurate binary fraction for the sampled clusters. In fact, for clusters with a larger MS spread, we identify binMS with mass ratio q$\gtrsim$0.6, while for those with a much smaller spread, we could also be including q$\gtrsim$0.4 or even lower. A much more accurate assessment, complemented by a differential reddening correction \citep{pancino24}, would be required to estimate accurately the fraction of MS binaries with different mass ratios, or to provide a homogeneous sample with q above a fixed value. Moreover, an accurate study of stellar crowding would be required to estimate the number of chance superpositions or blends \citep{milone12}, which is beyond the scope of the present paper.

Additionally, the method of determining the binary fraction from photometric sequences \citep[first proposed by][]{kahler99} presents two problems. The first is that, below a certain q, binaries land on the MS itself and it is difficult to disentangle them from normal MS stars. For this reason, it is necessary to make assumptions on the q distribution to obtain a binary fraction. Across different field and cluster samples, the q distribution has been found to be flat \citep{raghavan10,alexander25}, peaking at q$\simeq$0.23 \citep{duquennoy91}, peaking at q$\simeq$1 \citep{fisher05}, and multi-peak \citep{moe17}. The second problem is that many types of binaries and binary byproducts, including stars out of equilibrium or in particular configurations, also hide within the MS (such as BL). At the same time, many binaries and their byproducts lie outside of the canonical sequences (BSS, subdwarfs, and so on) and are completely excluded if only the MS and binMS are considered. 

Thanks to our large compilation of known or suspected binaries from the literature, we are in a position to provide at least some clues on what might be a reasonable q distribution in open clusters. For the MS sample, we find 7\% of confirmed binaries, regardless on whether we include or not less certain ones as well ({\tt flag\_bin}\,=\,{\tt BIN:}). For the binMS sample,  the corresponding figure is 6\%. One important factor to consider is that, while it is impossible to correct such a compilation of different methods for detection bias, at the same time the MS and binMS samples should suffer from very similar detection biases. In other words, the discovery fractions reported above should in principle relate to each other in the same way as the actual binary fractions. Our values can thus be used to provide a rough estimate of the ratio between binaries with q$\gtrsim$0.6 and those with q$\lesssim$0.6, which is about 0.8. This could be a useful indication for future studies of binaries in open clusters. For example, if the q distribution was flat, we would obtain about 0.7.

%%%%%%%%%%%%%%%%%%%%%%%%%%%%%%%%%%%%%%%%%%%%%%%%%%

\subsection{Candidate blue lurkers}
\label{sec:bl}

Lower mass BSS, hidden in the normal MS of M\,67 and of other clusters, were discovered by means of their fast rotation \citep{leiner19}, which is interpreted as a sign of recent close binary interaction. Similarly to the case of BSS, evidence of close compact companions was found for a few BL so far \citep{jadhav19,nine23,panthi23}. According to \citet{leiner25}, at least one BL was originated in a triple system. Consequently, the operational definition of a BL has shifted in the literature, considering also the UV flux excess, as indicative of a compact companion, as a defining property. The lower masses of BL, compared to BSS, are thought to be caused by a less efficient mass transfer during the interaction \citep{sun24} or by a lower starting mass of the accretor. 

In this work, we define BL candidates as stars belonging to the MS or binMS sample of each cluster, with rotation higher than 3\,MAD with respect to other stars at a similar magnitude. Our main goal is in fact to decontaminate the MS sample from abnormally fast rotating stars, even if they could be other kinds of interacting binaries. We generally did not consider stars above the point in the MS where the internal structure of the star changes, i.e. the Kraft break \citep{kraft67}, especially in clusters with an eMSTO or a split MS. These stars have a large spread in the rotational distribution, which makes it difficult to clearly identify outliers. Therefore, our sample of BL candidates is dominated by low-mass stars, not all of which will be true BL. To this, we add the BL found in the literature \citep[][and Reggiani et al., in preparation]{nine23,jadhav19,jadhav23,jadhav24,jadhav26}. However in the case of NGC\,6940 and NGC\,752 the literature BL fall onto the eMSTO region and thus we did not consider them here. More in general, we did not flag as BL any star that was already flagged as eMSTO or splitMS. One notable property of our compilation of literature and new BL candidates is that they reach lower ages compared to most literature sources (6.8\,$\lesssim$\,log(age)\,$\lesssim$\,9.8\,dex).

The {\tt vbroad} distribution of our BL candidates peaks at about 50\,km\,s$^{-1}$, with a tail reaching 250\,km\,s$^{-1}$ and above (Fig.~\ref{fig:histbinDW}). About 20\% (23 over 113) of the candidate BL are flagged as binaries or suspected binaries, which is about twice the percentage of the MS sample (or the binMS one). They are mostly classified as SB1, with one SB2 and a few astrometric binaries. When the periods are available, they are mostly above 200\, days. Eight candidates are classified as RS\,CVn in {\em Gaia} DR3 (but not parametrised) and one is an eclipsing binary with a period of 0.9\,days in ASAS-SN and \citet{gavras23}. Many of the candidates were flagged as rotational variables in the literature.  

%%%%%%%%%%%%%%%%%%%%%%%%%%%%%%%%%%%%%%%%%%%%%%%%%%

\subsection{Additional figures}
\label{app:fig}

\begin{figure*}[!h]
    \centering
    \includegraphics[width=\textwidth]{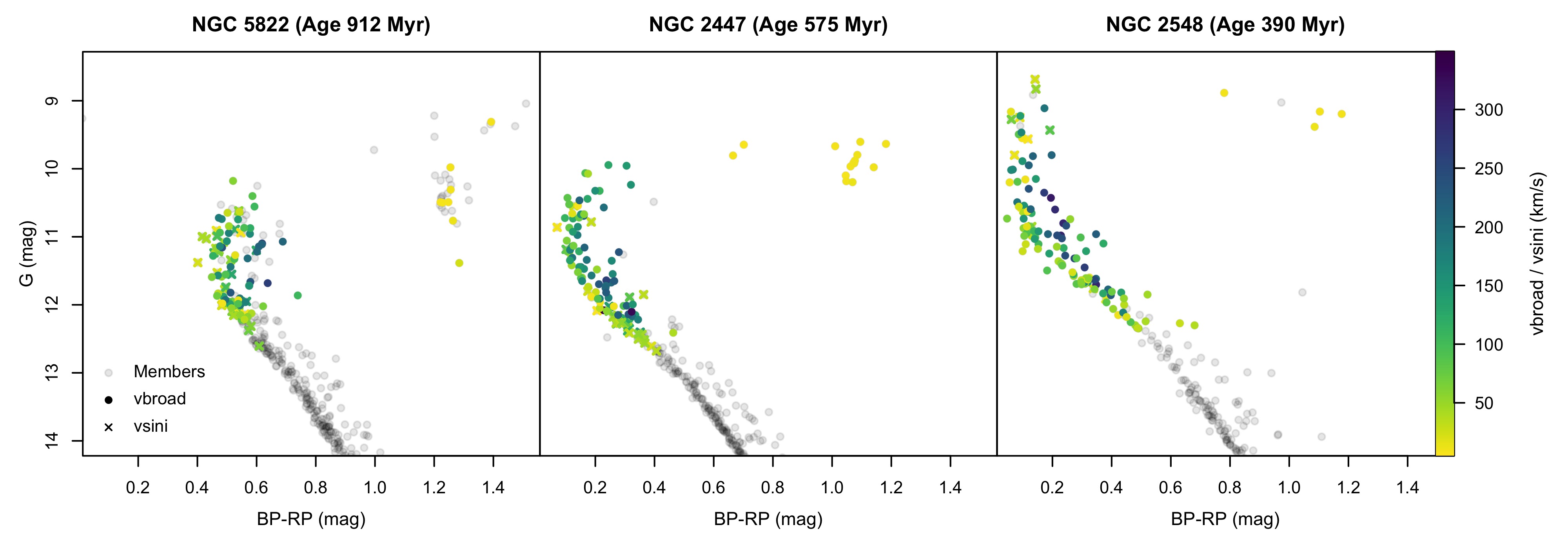}
    \includegraphics[width=\textwidth]{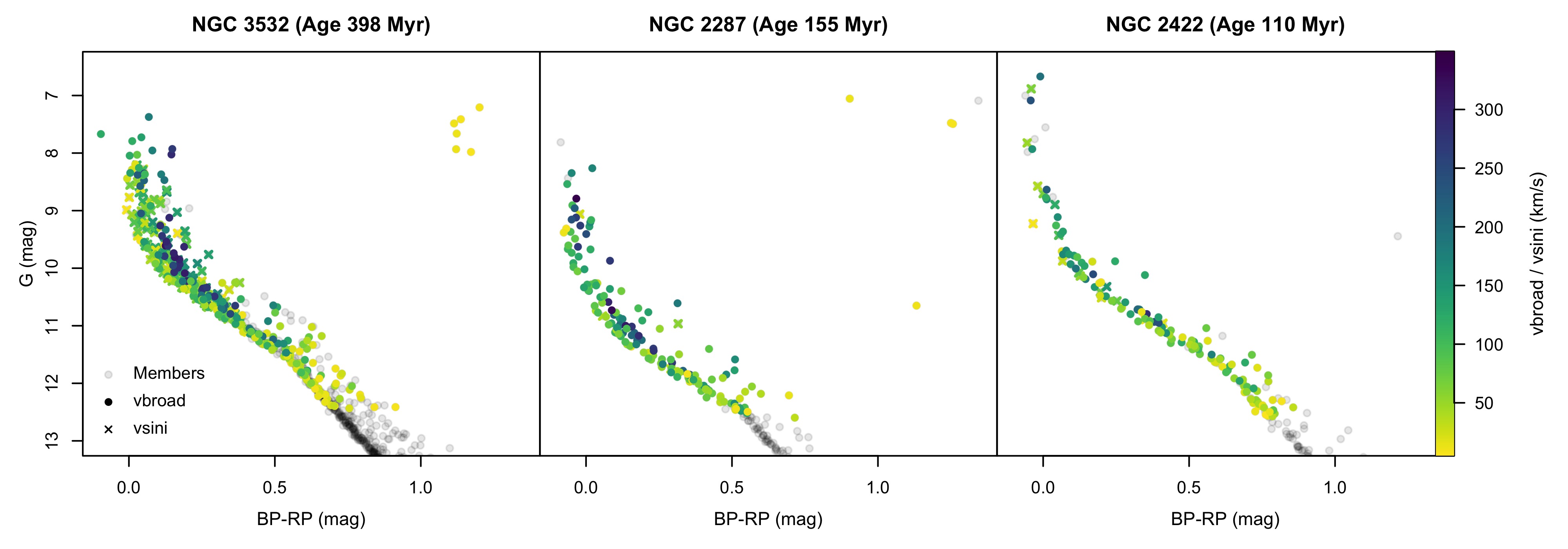}
    \caption{Examples of clusters with an eMSTO (top panels) or with a split MS (bottom panels), sorted by decreasing age. The whole cluster sample is plotted in grey in the background, while stars having a {\tt vbroad} estimate are plotted as colored circles and those having a {\tt vsini} estimate are plotted as colored crosses.}
   \label{fig:emsto}
\end{figure*}

We place here additional figures with examples of sample selections and specific clusters, relevant to Sect.~\ref{sec:res}. In particular, Fig.~\ref{fig:binMS} shows some examples of the separation between the MS and binMS samples, while Fig.~\ref{fig:emsto} presents a selection of clusters with clean eMSTO and split MS.

% --------------------------------------------
% --------------------------------------------

\end{document}